\documentclass[12pt]{iopart}
\usepackage{iopams}  
\usepackage{lscape}
\usepackage{graphicx}
\usepackage{threeparttable}
\usepackage{multirow}
\usepackage{color}
\newcommand{\mnras}{MNRAS}
\newcommand{\aap}{Astronomy and Astrophysics}
\newcommand{\apjs}{ApJS}
\newcommand{\apj}{ApJ}

\newcommand{\prd}{Phys. Rev. D}

\newcommand{\nat}{Nature}
\newcommand{\aone}{\alpha_1}
\newcommand{\atwo}{\alpha_2}
\newcommand{\athree}{\alpha_3}
\newcommand{\ai}{\alpha_i}
\newcommand{\haone}{\hat{\alpha}_1}
\newcommand{\hatwo}{\hat{\alpha}_2}
\newcommand{\hathree}{\hat{\alpha}_3}
\newcommand{\hai}{\hat{\alpha}_i}
\newcommand{\psrb}{PSR~B1937+21}
\newcommand{\psrj}{PSR~J1744$-$1134}
\newcommand{\psrs}{PSRs~B1937+21 and J1744$-$1134}

\begin{document}

\title[A new $\hatwo$ limit from solitary pulsars]{A new limit on
  local Lorentz invariance violation of gravity from solitary pulsars}

\author{Lijing Shao$^{1,2}$, R. Nicolas Caballero$^{1}$, Michael
  Kramer$^{1,3}$, Norbert Wex$^{1}$, David J. Champion$^{1}$, Axel
  Jessner$^{1}$}

\address{$^1$\,Max-Planck-Institut f\"{u}r
  Radioastronomie, Auf dem H\"{u}gel 69, D-53121 Bonn, Germany}
\address{$^2$\,School of Physics,
  Peking University, Beijing 100871, China}
\address{$^3$\,Jodrell Bank Centre for Astrophysics, School of Physics
  and Astronomy, \\ The University of Manchester, M13 9PL, UK}
  
\eads{\mailto{lshao@pku.edu.cn (LS)}
}

\begin{abstract}

Gravitational preferred frame effects are generally predicted by
alternative theories that exhibit an isotropic violation of local
Lorentz invariance of gravity. They are described by three parameters
in the parametrized post-Newtonian formalism.  One of their
strong-field generalizations, $\hatwo$, induces a precession of a
pulsar's spin around its movement direction with respect to the
preferred frame. We constrain $\hatwo$ by using the non-detection of
such a precession using the characteristics of the pulse profile. In
our analysis we use a large number of observations from the 100-m
Effelsberg radio telescope, which cover a time span of approximately
15 years.  By combining data from two solitary millisecond pulsars,
\psrs{}, we get a limit of $|\hatwo|<1.6\times10^{-9}$ at 95\%
confidence level, which is more than two orders of magnitude better
than its best weak-field counterpart from the Solar system.

\end{abstract}


\maketitle

\section{Introduction}
\label{sec:intro}

There are many models and test frameworks involving Lorentz violation
in the gravitational sector, such as the vector-tensor theory
in~\cite{wn72}, TeVeS gravity~\cite{bek04,sag09}, Einstein-\AE{ther}
theory~\cite{jm01}, Ho{\v r}ava-Lifshitz gravity~\cite{hor09,bps11},
and the standard model extension (SME) of gravity~\cite{kos04,bk06}. A
preferred frame, possibly associated with the distribution of matter
in the universe, may result, if the Lorentz violation is isotropic in
a specific frame.

The existence of a preferred frame would induce various preferred
frame effects (PFEs) that can be probed through different physical
observables.  In the parametrized post-Newtonian (PPN)
formalism~\cite{wn72,wil93}, PFEs are characterized by three
parameters, $\aone$, $\atwo$ and $\athree$. In Einstein's general
relativity (GR), $\aone=\atwo=\athree=0$. Their strong-field
generalizations are denoted as $\haone$, $\hatwo$, and $\hathree$, and
again in GR, $\haone=\hatwo=\hathree=0$. Nevertheless, one can have
non-zero values of these parameters in alternative gravity
theories~\cite{wil93,wil06}.

Observational implications of PFEs have been studied by several
authors using different methods, $\ai$ (as well as $\hai$; $i=1,2,3$)
are constrained to high precision from geophysics, Solar system, and
pulsar timing experiments~\cite{nw72,de92,sfl+05,wk07,sw12}.  We
briefly present the best limits of $\ai$ (or $\hai$) below.
\begin{itemize}
\item
  Currently, the best limit on $\haone$ comes from the orbital
  dynamics of the binary pulsar PSR~J1738+0333~\cite{fwe+12}, which
  gives a robust limit~\cite{sw12},
  \begin{equation}\label{eq:a1psr}
    \haone = -0.4^{+3.7}_{-3.1}\times10^{-5}\,, \quad \mbox{(95\%
      CL)}.
  \end{equation}
\item
  The best limit on $\atwo$ comes from the alignment of the Sun's spin
  with the orbital angular momentum of the Solar system~\cite{nor87}
  (note, $\atwo^{\rm Nordtvedt} = \frac{1}{2}\atwo$), which gives
  \begin{equation}\label{eq:a2solar}
    |\atwo|<2.4\times10^{-7}\,.
  \end{equation}
\item
  The best limit on $\hathree$ comes from the orbital dynamics of the
  statistical combination of a set of binary pulsars~\cite{sfl+05},
  which gives a probabilistic limit,
  \begin{equation}
    |\hathree| < 4.0 \times 10^{-20}\,, \quad \mbox{(95\% CL)}.
  \end{equation}
\end{itemize}
As shown above, except for $\atwo$, pulsar timing observations have
provided better limits than those from Solar system experiments. Here,
however, one has to keep in mind, that pulsars are also sensitive to
strong-field deviations, which do not occur in the weak-field regime
of the Solar system. We note that, in general the preferred frame is
not specified.  The most natural option from a cosmological
perspective is the frame where the cosmic microwave background (CMB)
radiation is isotropic. The results discussed in this paper correspond
to such a frame, however, see e.g. \cite{wk07,skmb08,sw12} for other
preferred frames.

Since the first pulsar discovery in 1967~\cite{hbp+68} more than two
thousand pulsars have been discovered and studied with radio, X-ray
and $\gamma$-ray observations \cite{mhth05}. These celestial objects
are intriguing in multiple aspects, e.g., some of them show a
long-term rotational stability similar to the stability of atomic
clocks~\cite{hcm+12}, their high interior density exceeds that of
nuclear matter, and their high magnetic field is comparable to or even
exceeds the quantum critical value (see references in~\cite{lk05}).
In particular, one of the most important contributions of pulsars is
their unique r\^ole in tests of gravity theories, especially in the
investigation of strong-field deviations from GR. To highlight some
great achievements: i) the Hulse-Taylor pulsar provided the first
evidence for the existence of gravitational waves \cite{tfm79,tw82};
ii) the double pulsar provided the most accurate tests of GR in the
strong-field regime, up to a precision of 0.05\%~\cite{ksm+06a}; iii)
pulsar white dwarf systems provided the most stringent tests on the
scalar-tensor theories~\cite{fwe+12,afw+13}.  In this paper, we report
a new limit on the (strong-field) PPN parameter $\hatwo$ from solitary
millisecond pulsars (MSPs), which surpasses its current best
weak-field counterpart from the Solar system~\cite{nor87} by more than
two orders of magnitude.

The paper is organized as follows. In the next section, we introduce
the $\atwo$ (and $\hatwo$) parameter and its effect on the spin vector
of solitary pulsars, which lays the principle of the test. In section
\ref{sec:pulse}, we present our two solitary pulsars, \psrs{}, and the
analysis of a large number of observations spanning about 15 years
that were obtained from the 100-m Effelsberg radio telescope. Then in
section~\ref{sec:a2}, by using the non-detection of profile variation,
we set a greatly improved constraint of $|\hatwo| < 1.6\times10^{-9}$
at 95\% confidence level (CL). Section~\ref{sec:con} discusses the
relevance of our new limit, and briefly summarizes the
paper. Throughout the paper, we use boldface letters to represent
vectors, and put ``hat'' onto them to indicate their corresponding
unit vectors.  Strong-field generalizations of PPN parameters are
distinguished explicitly by adding a ``hat'' onto their corresponding
weak-field counterparts.

\section{Preferred frame effects and $\hatwo$-induced spin precession}
\label{sec:pfe}

\begin{figure}
  \begin{center}
    \includegraphics[width=11cm]{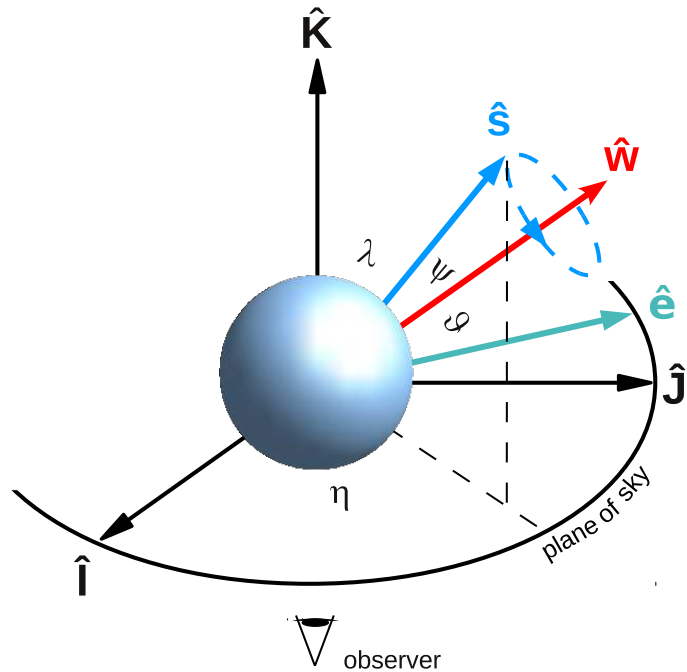}
  \end{center}
\caption{\label{fig:geometry} Angle notations and the $\hatwo$-induced
  precession of the pulsar spin axis $\hat{\bf s}$ around $\hat{\bf
    w}$, the movement direction of pulsar with respect to the
  preferred frame (see text). The coordinate system ($\hat{\bf I},
  \hat{\bf J}, \hat{\bf K}$) is defined with $\hat{\bf I}$ pointing to
  east, $\hat{\bf J}$ pointing to the north celestial pole, and
  $\hat{\bf K}$ pointing along the line of sight. The unit vector
  $\hat{\bf e} \equiv \hat{\bf K} \times \hat{\bf s} / | \hat{\bf K}
  \times \hat{\bf s}|$ is in the sky plane.}
\end{figure}

Let us first summarize some key theoretical ingredients for the
$\atwo$ test. The $\atwo$-related many-body post-Newtonian Lagrangian
term reads \cite{nor87,de92},
\begin{equation}\label{eq:la2}
 L_{\atwo} = \frac{\atwo}{4} \sum_{i\neq j} \frac{{G}m_im_j}{c^2r_{ij}}
 \left[ (\mathbf{v}_i^0 \cdot \mathbf{v}_j^0) -
   (\hat{\mathbf{n}}_{ij}\cdot\mathbf{v}_i^0)
   (\hat{\mathbf{n}}_{ij}\cdot\mathbf{v}_j^0) \right] \,,
\end{equation}
where $\mathbf{v}^0_i$ is the velocity of body $i$ with respect to the
preferred frame, $r_{ij}$ is the coordinate separation of objects $i$
and $j$, and $\hat{\mathbf{n}}_{ij} \equiv
(\mathbf{r}_i-\mathbf{r}_j)/r_{ij}$. The velocity of the
center-of-mass of the many-body system with respect to the preferred
frame we denote by ${\bf w}$. Nordtvedt showed that, as a result of
(\ref{eq:la2}), the spin axis of a massive body precesses around
$\mathbf{w}$.  The precession has an angular velocity~\cite{nor87},
\begin{equation}\label{eq:omegaprec}
\Omega^{\rm prec} = -\frac{\atwo}{2} \left( \frac{2\pi}{P}\right)
\left( \frac{w}{c}\right)^2 \cos \psi \,,
\end{equation}
where $P$ is the spin period of the body's rotation, $\psi$ is the
angle between $\mathbf{w}$ and the spin direction $\hat{\bf s}$ (see
figure~\ref{fig:geometry} for angles and an illustration of the
precession), and $w\equiv|{\bf w}|$. A similar consequence was also
found in the orbital dynamics of a binary system~(see (24)
in~\cite{sw12}), where a (strong-field) $\hatwo$ induces a precession
of the orbital angular momentum around ${\bf w}$ for a
small-eccentricity binary.

As mentioned before, Nordtvedt~\cite{nor87} used the current alignment
of the Sun's spin with the orbital angular momentum of the Solar
system, to limit such a precession. His limit~(\ref{eq:a2solar}) has
remained the best limit of $\atwo$ for more than a quarter of a
century. The crucial assumption inherent is that the Sun's spin {\it
  was} aligned with the Solar system angular momentum five billion
years ago when the Sun was born. A weaker but more robust limit on
$\atwo$ comes from a long-term project called lunar laser ranging
(LLR), which gives~\cite{mwt08}
\begin{equation}\label{eq:a2llr}
\atwo =(1.8\pm5.0)\times10^{-5}\,,\quad \mbox{(95\% CL)},
\end{equation}
from an analysis of 35 years of data.  It is two orders of magnitude
weaker than that of (\ref{eq:a2solar}). The best limit in the strong
field is from pulsar timing experiments on pulsar binaries
PSRs~J1012+5307 and J1738+0333~\cite{sw12},
\begin{equation}\label{eq:a2bnrypsr}
  |\hatwo| < 1.8 \times 10^{-4} \,, \quad \mbox{(95\% CL)} .
\end{equation}

The remarkable limit~(\ref{eq:a2solar}) obtained by
Nordtvedt~\cite{nor87} benefited enormously from a long baseline of
time of approximately five billion years.  However, as we can see from
(\ref{eq:omegaprec}), one can also take advantage of the short spin
period of MSPs to achieve a tight constraint. This method was
originally suggested in~\cite{nor87} shortly after the discovery of
the first millisecond pulsar. We present the first detailed analysis
in this direction.

With a non-vanishing $\hatwo$, a spinning pulsar would precess around
its ``absolute'' velocity, ${\bf w}$, with an angular velocity
(\ref{eq:omegaprec}).  As a result of the precession, the angle,
$\lambda$, between the pulsar spin axis and our line of sight changes
with time (see figure~\ref{fig:geometry}), so that different portions
of the pulsar emission beam are observed at different epochs.
Consequently, one expects to detect characteristic changes in the
measured pulse profile as a function of time. For solitary pulsars,
from pure geometrical consideration we have (see also (2)
in~\cite{bai88}),
\begin{equation}\label{eq:lambdadot}
 \frac{{\rm d}\lambda}{{\rm d}t} = \Omega^{\rm prec} \, \hat{\bf w} \cdot
 \left( \frac{ \hat{\bf K} \times \hat{\bf s} } { | \hat{\bf K} \times
   \hat{\bf s}|} \right) \equiv \Omega^{\rm prec} \cos \vartheta \,,
\end{equation}
where $\vartheta$ is the angle between $\hat{\bf w} \equiv {\bf w}/w$
and $\hat{\bf e} \equiv \hat{\bf K} \times \hat{\bf s} / | \hat{\bf K}
\times \hat{\bf s}|$. The unit vector, $\hat{\bf e}$, gives the line
of nodes associated with the intersection of the sky plane and the
equatorial plane of the pulsar (see figure~\ref{fig:geometry}).

Current observational technologies are already sensitive enough to
detect such a change, if it exists. Indeed, similar changes in pulsar
profiles have been observed before, albeit under the influence of
geodetic precession, e.g.~for, PSR~B1913+16, PSR~B1534+21,
PSR~J1141$-$6545, and
PSR~J0737$-$3039B~\cite{wrt89,kra98,sta04,mks+10,pmk+10}.  Geodetic
precession occurs in binary pulsars due to the curvature of spacetime
near gravitating bodies, where the proper reference frame of a freely
falling object suffers a precession with respect to a distant
observer.  The caused pulse profile changes manifested themselves in
various forms~\cite{dr75}, such as changes in the amplitude ratio or
separation of two pulse components~\cite{wrt89,kra98}, the shape of
the characteristic swing of the linear polarization~\cite{sta04}, or
the absolute value of the position angle~\cite{mks+10}.

\begin{figure}
    \hspace{4cm}
  \begin{center}
    \includegraphics[width=14cm]{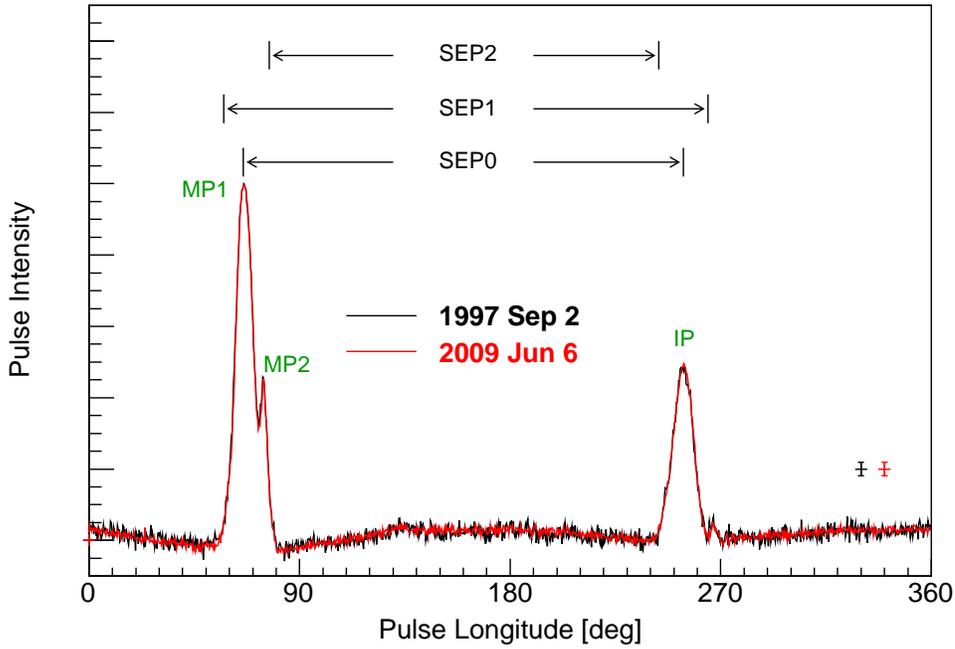}
  \end{center}
\caption{\label{fig:B1937_pulse} Comparison of two pulse profiles of
  \psrb{} obtained at two different epochs --- the black one was
  obtained on September 2, 1997, while the red one was obtained on
  June 6, 2009. The main peak is aligned and scaled to have the same
  intensity. Uncertainties in pulse profiles are illustrated at the
  right bottom corner. Notations used in figure~\ref{fig:B1937_width}
  include: the first component of the main-pulse ({\tt MP1}), the
  second component of the main-pulse ({\tt MP2}), the interpulse ({\tt
    IP}), the separation between {\tt MP1} and {\tt IP} ({\tt SEP0}),
  the separation between leading {\tt MP1} and trailing {\tt IP} ({\tt
    SEP1}), and the separation between trailing {\tt MP2} and leading
  {\tt IP} ({\tt SEP2}).}
\end{figure}

\begin{figure}
    \hspace{4cm}
  \begin{center}
    \includegraphics[width=14cm]{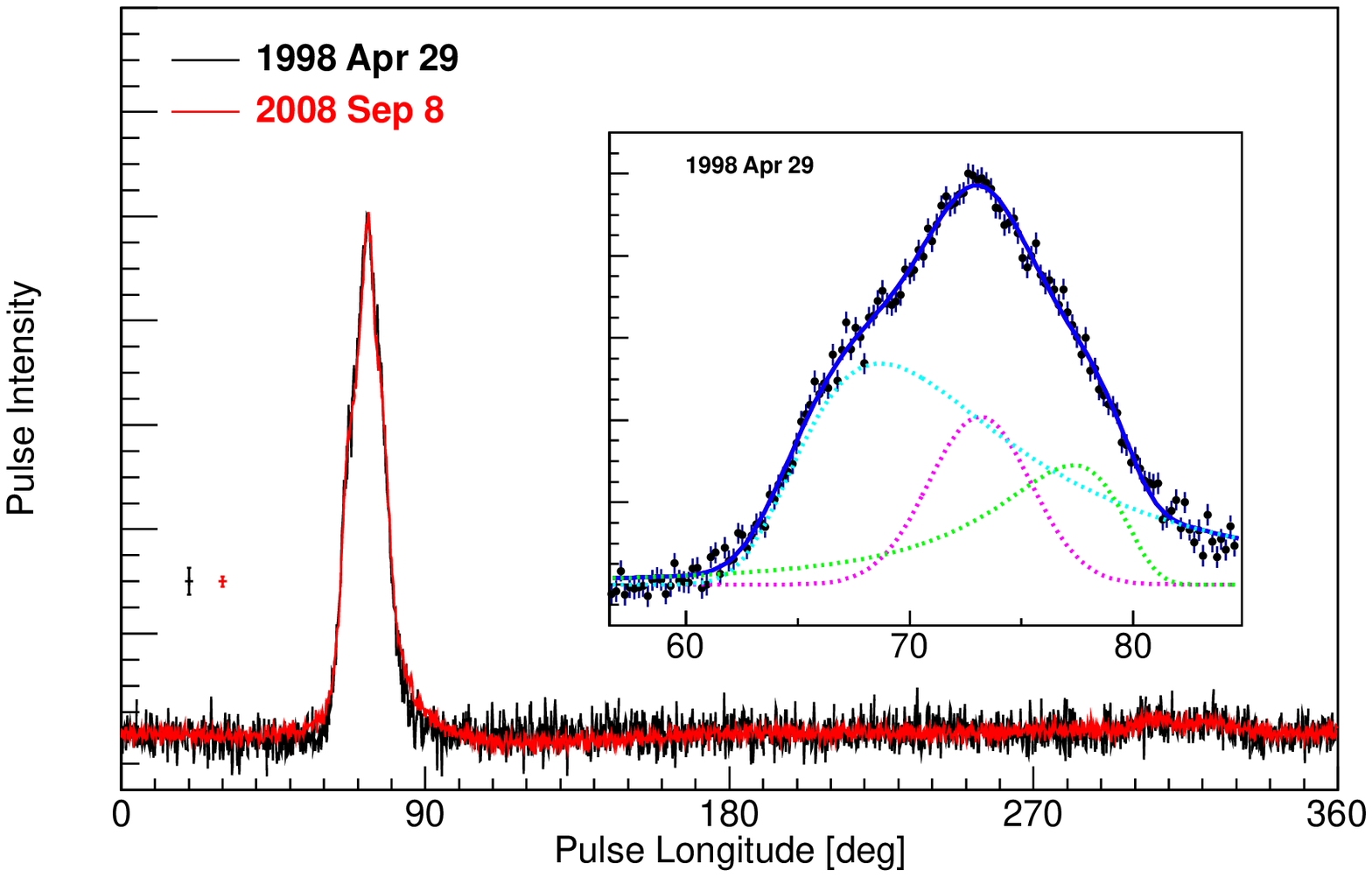}
  \end{center}
\caption{\label{fig:J1744_pulse} A comparison of two pulse profiles of
  \psrj{} obtained during two different epochs --- the black one was
  obtained on April 29, 1998, while the red one was obtained on
  September 8, 2008. The peak is aligned and scaled to have the same
  intensity. Uncertainties in pulse profiles are illustrated at the
  left bottom corner. Inset shows the zoom-in of the main pulse
  (corresponding to the black profile in the main figure), and it also
  shows our analytical fitting to the pulse and the corresponding
  three components (see text).}
\end{figure}

For our purpose, to avoid complications due to spin-orbit coupling, we
choose solitary pulsars to limit $\hatwo$.  In our solitary pulsars
below, if there exists an $\hatwo$-induced precession, we are also
expected to observe one or several of the above mentioned changes in
the pulse profile. On the other hand, if we do not see any changes in
the observations, we can constrain $\hatwo$. As an example of such a
non-detection, in figure \ref{fig:B1937_pulse} we plot two pulse
profiles of \psrb{} obtained at different epochs. One was obtained on
September 2, 1997, while the other was obtained on June 6,
2009. Details of the used observing system will be given given in
Section 3. Instrumental effects are responsible for the ``dips''
around the pulse.  We have not removed these effects since we only use
data from one backend and the dips remain unchanged in time and do not
introduce any temporal variation in the profiles.  We can immediately
see from figure \ref{fig:B1937_pulse} that within noise, there is no
visible change in the pulse profile for this pulsar over more than ten
years. The two profiles are chosen solely based on a large time
separation and a low level of noise, so that no bias is introduced.  A
similar overlap of two pulse profiles for the other pulsar in our
test, \psrj{}, is shown in figure~\ref{fig:J1744_pulse}. The profiles
were obtained on April 29, 1998 and September 8, 2008. There exists no
visible difference within noise level.  These two solitary pulsars are
selected from the known population of millisecond pulsars, based on
their figure of merit for the $\hatwo$ test. The figure of merit is
roughly $P^{-1} T_{\rm obs}^{3/2}$ where $T_{\rm obs}$ is the
observational time span.  We also require the pulsars to have proper
motion measurements from pulsar timing and geometry information from
the combination of radio and $\gamma$-ray observations. The reason for
the figure of merit and these requirements will become clear later.

To achieve a quantitive constraint, we need to relate the change in
$\lambda$ with that of a profile. One can confidently consider the
pulse profile as a cross-sectional cut through the pulsar's emission
beam \cite{lk05}. To quantify the (non-)change of pulsar geometry from
a pulse profile, we should introduce a basic emission model. We use
the simplest geometrical cone model~\cite{ggr84}, which only assumes
that radio beam is centred on the magnetic axis, causing the
``lighthouse'' effect of a pulsar as it rotates around the spin
axis. This approximation avoids most of the model dependent aspects of
pulsar emission, and sufficiently reproduces the basic features of the
two solitary pulsars we are using here. We note in passing that the
limit on the pulsar spin precession does not depend on this assumption
significantly, as shown in the geodetic precession analysis for
PSR~J0737$-$3039A~\cite{mkp+05}.  The latter is also a non-detection
case, where authors showed that a non-zero ellipticity for the
radiation beam gives no significantly improved fits to the data, and a
circular beam describes the data equally~\cite{mkp+05}.

From the cone model, it was shown from geometry~\cite{ggr84,lk05},
\begin{equation}
\sin^2 \left( \frac{W}{4} \right) = \frac{ \sin^2(\rho/2) -
  \sin^2(\beta/2)}{\sin(\alpha+\beta)\sin\alpha} \,,
\end{equation}
where $W$ is the width of the pulse, $\alpha$ is the angle between
$\hat{\bf s}$ and the magnetic axis, $\beta \equiv 180^\circ
-\lambda-\alpha$ is the impact angle, and $\rho$ is the semi-angle of
the opening radiating region (for details, see~\cite{lk05} and
references therein).

Adopting a plausible assumption that the radiation property ($\alpha$
and $\rho$) has no change during the observational span $\sim 15$
years \cite{lk05}, i.e. $\rmd\alpha/\rmd t = \rmd\rho/\rmd t = 0$, we
can relate the change in $\lambda$ with the change in the pulse width
(see also (4) in~\cite{cwb90}),
\begin{equation}\label{eq:w50dot}
\frac{\rmd\lambda}{\rmd t} = \frac{1}{2}\,\frac{ \sin (W/2)}{ \cot
  \lambda \cos (W/2) + \cot\alpha} \, \frac{\rmd
  {W}}{\rmd t} \equiv {\cal A} \, \frac{\rmd
  {W}}{\rmd t}\,,
\end{equation}
where ${\cal A} \equiv { \sin (W/2)}/ [2\cot \lambda \cos (W/2)
  +2\cot\alpha]$.  Hereafter we use the width at 50\% intensity level,
$W_{50}$, as a proxy of $W$.\footnote{Other choices, like the width at
  a 10\%-level do not change the result in the following
  significantly.}  Now we can quantify the (non-)change of the pulsar
orientation with respect to the Earth through the (non-)change in the
pulse width. In the next section, detailed constraints on the
(non-)change of pulse width are derived, which are used to put a limit
on $\hatwo$ in section~\ref{sec:a2}.

\section{Observations and pulse profile analysis}
\label{sec:pulse}

In this section, we present our observations of two solitary pulsars
with the 100-m Effelsberg telescope and our detailed analysis of pulse
profiles.

All data were obtained with the 100-m Effelsberg radio telescope,
operated by the Max-Planck-Institut f\"ur Radioastronomie, Bonn,
Germany. The observations are part of the pulsar timing program (see
e.g.~\cite{fhb+10,hlj+11}). In order to examine the profile stability
over time, it is important to use data obtained with as few changes in
the observing system as possible.  The major components of the system
are the telescope, the receiver (frontend) and the data processor
(backend).

The observations span from September 1997 for \psrb{} and January 1997
for \psrj{}, to the present. Receiver systems operating at a frequency
around 1400\,MHz were used, being sensitive to two orthogonally
left-hand and right-hand circularly polarised signals. The frequency
of the signals was mixed to baseband and fed into a data acquisition
system known as the Effelsberg-Berkeley Pulsar Processor
(EBPP)~\cite{bdz+97}.  The EBPP is a coherent dedispersion backend
which means that it completely removes the signal dispersion effect of
the interstellar medium (ISM) caused by free electrons along the line
of sight. If left uncorrected this causes an apparent broadening of
the pulse.  The time resolution of our equipment was 1.4\,$\mu$s for
\psrb{} and 0.6\,$\mu$s for \psrj{}.  In the backend the signal from
the channels is directed to the dedisperser boards where online
coherent dedispersion takes place according to the recorded dispersion
measure (DM)\footnote{DM is defined as the integrated column density
  of free electrons in the ISM along the line of sight, ${\rm DM}
  \equiv \int n_e \rmd l$~\cite{lk05}.}. The output signals are folded
using the topocentric pulse period (i.e. individual pulses are
phase-aligned and added), and are later integrated in phase.  The EBPP
is the longest-running coherent dedispersion backend dedicated to high
precision pulsar timing, making its database uniquely suited this
work.  The total bandwidth of the EBPP is dependent on the source's
dispersion smearing at the observing frequency and has a maximum value
of 112\,MHz. The observational bandwidth for \psrb{} is 44\,MHz, while
for \psrj{}, all data have 112\,MHz of bandwidth apart from the first
two observations in January 1997 which have 56\,MHz of bandwidth.  The
frontend of the telescope changed once in July 2009, resulting in a
change of the central frequency from 1410\,MHz to 1360\,MHz. Frequency
evolution of the profile is very small for MSPs but we quantified the
change in profiles in section~\ref{sec:psrb}.

The data were reduced using the {\tt PSRCHIVE}
package~\cite{hsm04}. Each profile we use is a $\sim30$ minute
integration. This is achieved by adding shorter integrations made with
no more than one hour separation. Throughout the pulse profile fitting
analysis discussed in the following, we use the off-pulse
root-mean-square as the profile's flux uncertainty.

\subsection{\psrb{}}
\label{sec:psrb}

\begin{table*}
  \caption{Relevant quantities of \psrs{} for the $\hatwo$ test. Most
    quantities are from pulsar timing~\cite{vbc+09}, while the
    orientation and radiation quantities ($\alpha$ and $\zeta$) were
    obtained from model fitting to radio and $\gamma$-ray lightcurves
    (\psrb{}~\cite{gjv+12} and \psrj{}~\cite{joh12}).  The
    Lutz--Kelker bias in the timing parallax was
    corrected~\cite{vlm10}.  The scattering timescales were calculated
    according to an empirical relationship in~\cite{bcc+04}, and
    listed for 1410\,MHz/1360\,MHz. The pulse width and its time
    derivative are from this work.  For \psrb{}, quantities for {\tt
      MP1} (left) and {\tt IP} (right) are both tabulated.
    Parenthesized numbers represent the $1$-$\sigma$ uncertainty in
    the last digits quoted. \label{tab:psr}}
\begin{center}
  \begin{tabular}{lccc}
    \br
    Pulsar & \multicolumn{2}{c}{B1937+21} & J1744$-$1134 \\
    \mr
    Discovery (year) & \multicolumn{2}{c}{1982~\cite{bkh+82}} &
    1997~\cite{bjb+97} \\
    Right Ascension, $\alpha$ (J2000) & \multicolumn{2}{c}{$19^{\rm
        h}39^{\rm m}38^{\rm s}\!.561297(2)$} & $17^{\rm h}44^{\rm
      m}29^{\rm s}\!.403209(4)$ \\ 
    Declination, $\delta$ (J2000) &
    \multicolumn{2}{c}{$+21^\circ34'59''\!.12950(4)$} &
    $-11^\circ34'54''\!.6606(2)$ \\ 
    Spin period, $P$ (ms) & \multicolumn{2}{c}{1.55780653910(3)} &
    4.074545940854022(8) \\
    Reference epoch for $\alpha$, $\delta$ and $P$ (MJD) &
    \multicolumn{2}{c}{54219} & 53742 \\
    Proper motion in $\alpha$, $\mu_\alpha$ (mas\,yr$^{-1}$) &
    \multicolumn{2}{c}{$0.072(1)$}  & 18.804(8) \\
    Proper motion in $\delta$, $\mu_\delta$ (mas\,yr$^{-1}$) &
    \multicolumn{2}{c}{$-0.415(2)$} & $-9.40(3)$ \\
    Parallax, $\pi$ (mas) & \multicolumn{2}{c}{$0.14^{+0.05}_{-0.03}$}
    & 2.4(1) \\ 
    Dispersion measure, DM (cm$^{-3}$\,pc) &
    \multicolumn{2}{c}{71.0227(5)} & 3.1380(3) \\
    Magnetic inclination, $\alpha$ (deg) & $75^{+8}_{-6}$
    & $105^{+6}_{-8}$ & $51^{+16}_{-19}$ \\
    Observer angle, $\zeta$ (deg) & \multicolumn{2}{c}{$80(3)$} &
    $85^{+3}_{-12}$ \\ 
     Scattering timescale, $\tau_{\rm s}$ (ns) & \multicolumn{2}{c}{826/949}
     & 0.20/0.23 \\
     Time span of data used in this work (MJD) &
     \multicolumn{2}{c}{50693--55725} & 50460--55962\\ 
    Pulse width at 50\% intensity, $W_{50}$ (deg) & 8.281(9) &
    10.245(17) & 12.53(3) \\ 
    Time derivative of $W_{50}$, $\rmd{W}_{50}/\rmd t$
    ($10^{-3}\,\mbox{deg}\,{\rm yr}^{-1}$) & $-3.2(34)$ & 3.5(66) &
    $1.3(72)$ \\ 
    Jump between two frequencies, $\Delta W_{50}$ (deg) & $0.12(3)$
    & 0.04(6) & --
    \\
    \br
  \end{tabular}
\end{center}
\end{table*}

\begin{figure}
  \begin{center}
    \includegraphics[width=15cm]{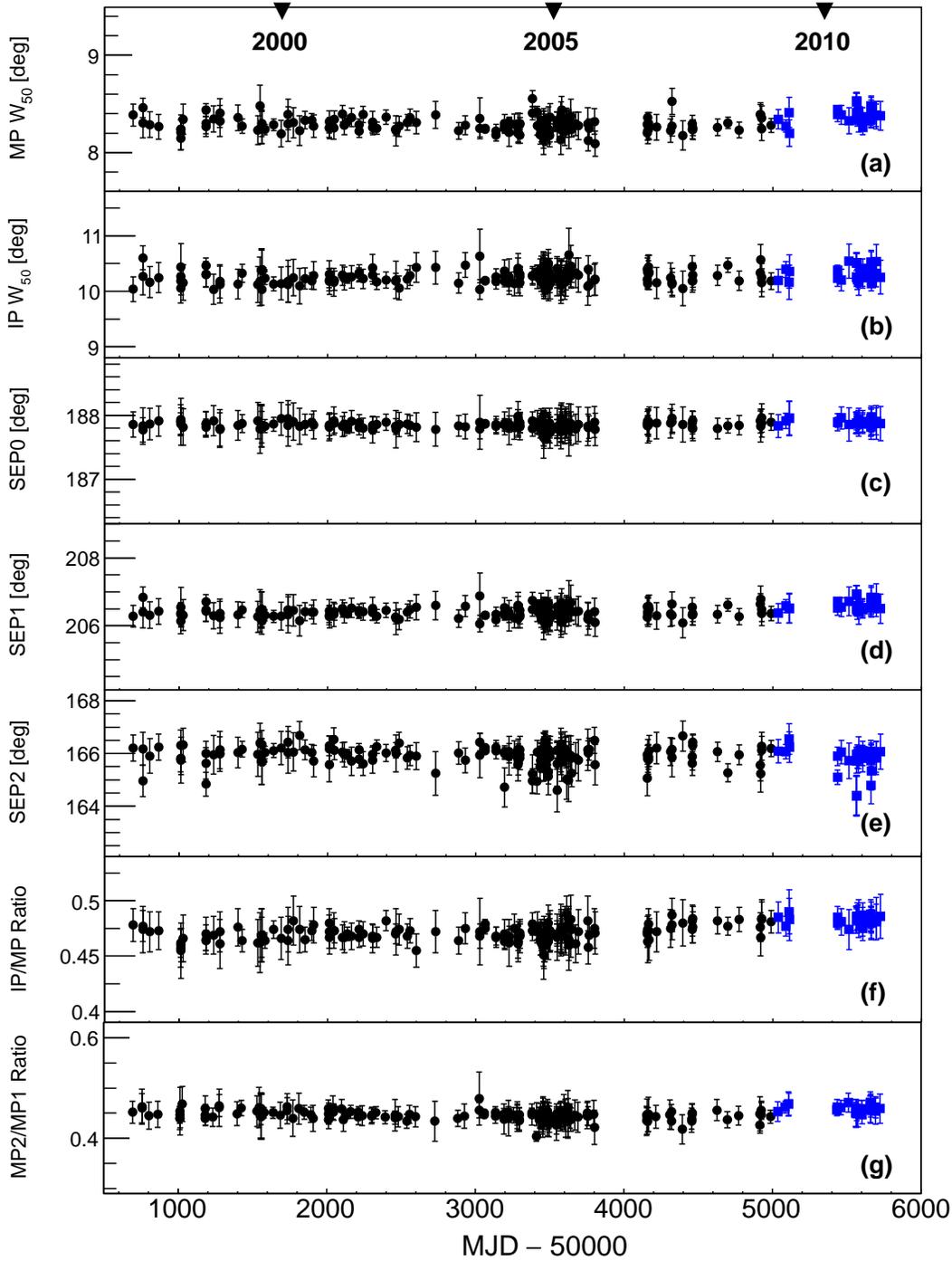}
  \end{center}
  \caption{\label{fig:B1937_width} Pulse profile characteristics of
    \psrb{}, as a function of modified Julian date (MJD); see
    figure~\ref{fig:B1937_pulse} for notations.  The amplitude ratios
    in (f--g) are measured from peak to peak.  Black circles are
    observations made at 1410\,MHz, while blue squares are
    observations made at 1360\,MHz.  Years of observations are
    indicated at the top of the figure.}
\end{figure}

\psrb{} (a.k.a PSR J1939+2134) was the first MSP discovered, with a
spin period of 1.56\,ms~\cite{bkh+82}. Because of its brightness and
later as a target of pulsar timing array (PTA)
projects~\cite{haa+10,fhb+10,hlj+11}, it has been observed frequently
since its discovery. \psrb{} shows a strong main-pulse ({\tt MP1})
with a second weaker component ({\tt MP2}) and a strong interpulse
({\tt IP}), see figure~\ref{fig:B1937_pulse} for illustrations. The
main-pulse and interpulse are separated by $\sim188^\circ$, hence they
may be produced by two opposite magnetic poles sweeping over the
Earth. Another possibility is that they are produced by a single
magnetic pole with a wide opening angle and a hollow cone emission
pattern.

In order to examine the profiles, we performed least-square fitting of
parabolas to the peaks of the three components. The simplicity of the
components' shapes allows good fits with a simple and symmetric
function, preserving linearity of the fitting procedure.  For each
component we obtained the peak intensity and its corresponding
longitude value as well as its $W_{50}$. We investigated the time
stability of the pulse profile using five different measurements of
widths and component separations (see figure~\ref{fig:B1937_pulse} for
definitions) and also two measurements of amplitude ratios of
different components. The results from fitting are plotted in
figure~\ref{fig:B1937_width} as a function of time.

\begin{figure}
  \begin{center}
    \includegraphics[width=18cm]{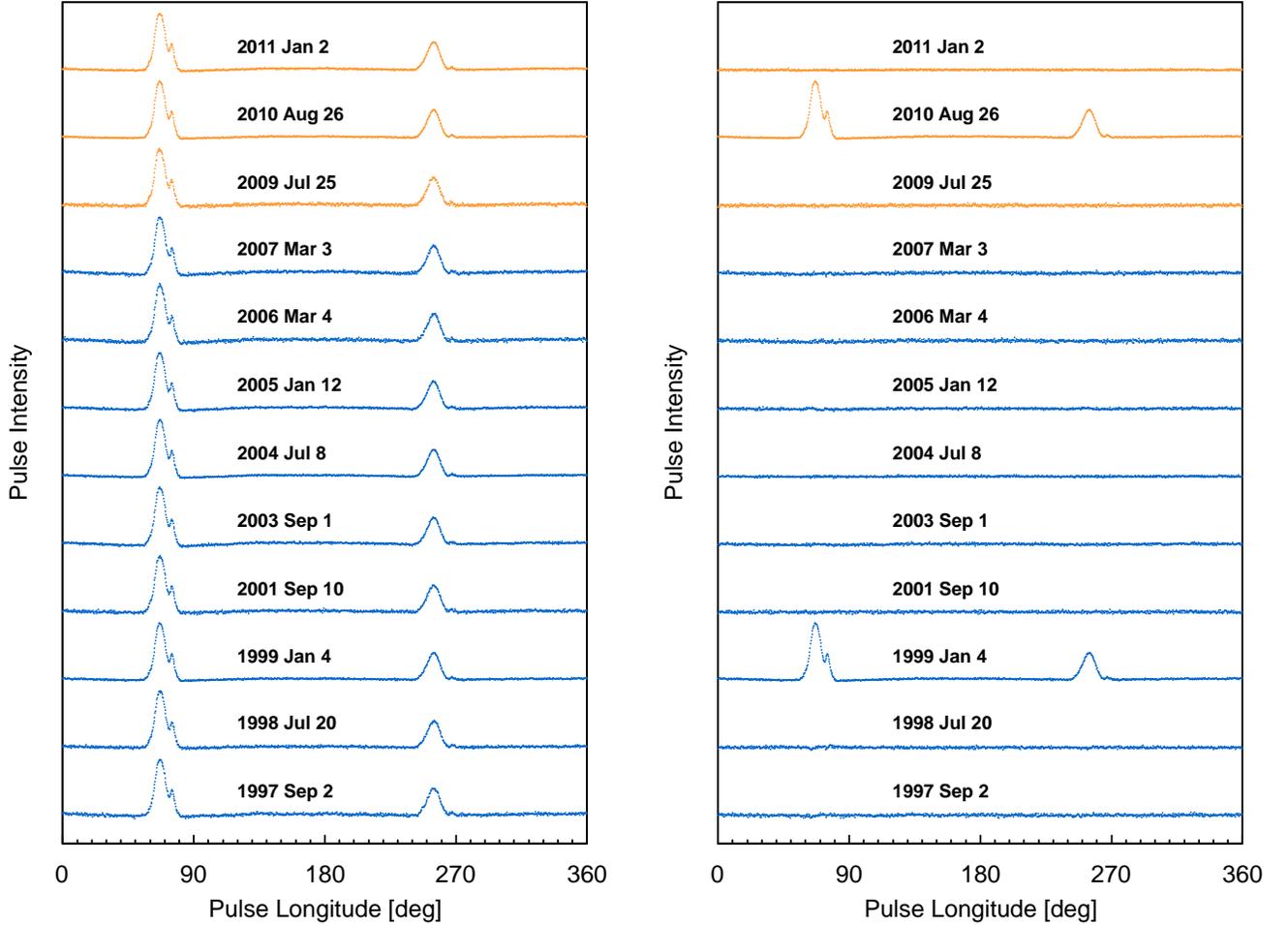}
  \end{center}
\caption{Integrated pulse profiles and difference profile residuals
  for \psrb{}, obtained from the 100-m Effelsberg radio telescope.
  Orange/Blue profiles are obtained from observations made with the
  central frequency 1360\,MHz/1410\,MHz.  {\em Left:} Twelve high S/N,
  aligned and normalized pulse profiles that are labeled with dates of
  the observation epochs. {\em Right:} A reference profile is
  subtracted from each of the other profiles of the same observational
  frequency. The pulse profiles taken on January 4, 1999 and August
  26, 2010 are used as reference profiles for two different
  frequencies; reference profiles are still plotted in the right
  panel.
\label{fig:B1937_diff}}
\end{figure}

One can see that the seven quantities characterizing the pulse profile
are very stable over the 15 years of observation.  We also plotted
twelve high signal to noise ratio (S/N) profiles in the left panel of
figure~\ref{fig:B1937_diff}. These profiles span almost uniformly the
whole observing period from 1997 to 2011. In the right panel of
figure~\ref{fig:B1937_diff}, the difference between profiles is
present after subtracting a reference profile. The highest S/N
reference profiles are chosen for each frequency --- one obtained on
January 4, 1999 for 1410\,MHz and one obtained on August 26, 2010 for
1360\,MHz. We can see from the residuals that no evolution over time
in the profile is visible.

Concerning the frequency dependence of $W_{50}$, in general normal
pulsars\footnote{Normal pulsar usually means a non-recycled pulsar
  with a spin period $P$ larger than 30\,ms, and a spindown rate
  $\dot{P} \sim 10^{-18}$ to $10^{-15}{\rm \,s\,s}^{-1}$; see
  references in~\cite{lk05} for details.} show a systematic increase
in pulse width when observed at lower frequencies \cite{lk05}, while
MSPs in general show very little evolution of pulse width with
frequency~\cite{kll+99}. Nevertheless, for both normal pulsars and
MSPs, profile evolution, in terms of peak intensities, width and
phase, was observed.  In figure~\ref{fig:B1937_width}, one can see
that there is no large difference between two frequencies.  However,
in our accurate data, we found that a subtle change between two
frequencies is needed. Therefore we fitted $W_{50}$'s of the
main-pulse and the interpulse with the following formula,
\begin{equation}\label{eq:linearfitting}
  W_{50}(t) = W_{50} + \frac{\rmd{W}_{50}}{\rmd t}\,t + 
  \Delta W_{50} \,\Theta (t-t_0) \,,
\end{equation}
where $\Delta W_{50}$ is the ``jump'' of width between measurements
made at 1360\,MHz and 1410\,MHz, $\Theta(t)$ is the Heaviside step
function, and $t_0$ is the time when observations were shifted from
1410\,MHz to 1360\,MHz.

We simulated $10^6$ sets of profile widths against time according to
the measurements and uncertainties in
figure~\ref{fig:B1937_width}\,(a), and then fitted simultaneously for
three parameters in (\ref{eq:linearfitting}) for each set of profile
widths. The fitted parameters were accumulated as histograms with
$10^6$ entries.  We read out the fitting results and their
uncertainties from the central values and the widths of these
histograms respectively.  They are tabulated in the last three rows in
table~\ref{tab:psr}, where uncertainties are rescaled by the square
root of the reduced fitting $\chi^2$, $\chi^2_{\rm red}$. The results
show the need of a tiny jump $\Delta W_{50} \simeq 0.1^\circ$ (about
one third of a bin in the EBPP profile data).

We investigated the possible origin of the jump by effects associated
with ISM.  For pulsars such as \psrb{} with high DM, $\Delta W_{50}$
may reflect the scattering effects from the irregularities in the ISM,
which usually produce a one-sided exponential tail for pulsar
profiles~\cite{lk05}.  The scattering timescales for the two
frequencies are listed in table~\ref{tab:psr}, and are based on an
empirical dependence on DM and frequency~\cite{bcc+04}. The broadening
in the pulse width from the intrinsic profile to profile at 1410\,MHz
is less than $0.02^\circ$ for \psrb{} from scattering, and the
difference between 1410\,MHz and 1360\,MHz is roughly $\left(\Delta
W_{50}\right)^{\rm scattering} \simeq 0.004^\circ$, hence
negligible. The empirical relation in~\cite{bcc+04} is poorly
constrained and may introduce overestimation or underestimation to an
amount of several times, however, after taking the uncertainties into
account, the effects from scattering are still too small to account
for the $\Delta W_{50}$ we obtained from fitting.  Another factor to
consider is temporal DM variation.  It is well known that the DM
varies with time and various efforts were made to measure these
variations systematically, see e.g.~\cite{yhc+07}.  If the DM at the
observatory is not properly updated, then the dedispersion would be
imperfect, leading to a broader pulse. We have calculated the width
difference between the two frequency bands for a DM value that
deviates as much as 0.05\,cm$^{-3}\,$pc from the correct value (it
would take decades without updating the DM value to get such a large
deviation for \psrb{}~\cite{rdb+06,yhc+07}), and we found that the
$\Delta W_{50}$ between the 1410\,MHz and 1360\,MHz bands are below a
few times $0.001^\circ$.  Therefore, the effects from DM variation are
also negligible.  In order to test for the possibility that $\Delta
W_{50}$ is caused by profile evolution we made use of data taken at
Effelsberg with {\tt Asterix}, a new backend that has run in parallel
with the EBPP since 2011.  {\tt Asterix} has a broader bandwidth of
200\,MHz.  The frequency range covers both frequency bands of the EBPP
backend. We selected the {\tt Asterix} frequency range accordingly to
emulate the EBPP characteristics and found a jump $\Delta W_{50}
\simeq 0.07^\circ\pm0.03^\circ$ between two frequencies, which is
consistent with the jump from the EBPP backend. Consequently, we
conclude that $\Delta W_{50}$ reflects an evolution of the pulsar
profile width with frequency\footnote{See also figure 13
  in~\cite{kll+99} for the evolution of pulse profile of \psrb{} with
  frequency. Note that in the relevant frequency bands, the dip
  between {\tt MP1} and {\tt MP2} gets deeper when frequency
  increases, consequently the width of {\tt MP1} gets narrower. This
  is also consistent with the $\Delta W_{50}$ measurement here.}.

The same Monte Carlo fitting analysis was also applied to the
interpulse (figure~\ref{fig:B1937_width}\,(b)); see
table~\ref{tab:psr} for the fitted results. We detected no changes in
the pulse width against time for the interpulse as well.

For both the main-pulse and the interpulse, we find no evidence of
changes in the pulse width over time.  We also performed hypothesis
test to test the necessity of a non-zero $\left({\rmd{W}_{50}}/{\rmd
  t}\right)$ for main-pulse and interpulse.  Our null hypothesis is
that, the inclusion of $\left({\rmd{W}_{50}}/{\rmd t}\right)$ does not
provide a significantly better fit.  $F$-tests give $p$-values of 0.22
and 0.31 for the null hypothesis of the main-pulse and interpulse
respectively, which clearly show that the inclusion of a non-zero
$\left({\rmd{W}_{50}}/{\rmd t}\right)$ does not provide a
significantly better fit to the data\footnote{The $p$-value from the
  test is the probability of obtaining a test statistic at least as
  extreme as the one that is actually observed, assuming that the null
  hypothesis is true~\cite{ptvf92}. In our cases, the test statistic
  is $F$ statistic which follows an $F$ distribution.}.

\subsection{\psrj{}}
\label{sec:psrj}

\begin{figure}
  \begin{center}
    \includegraphics[width=17cm]{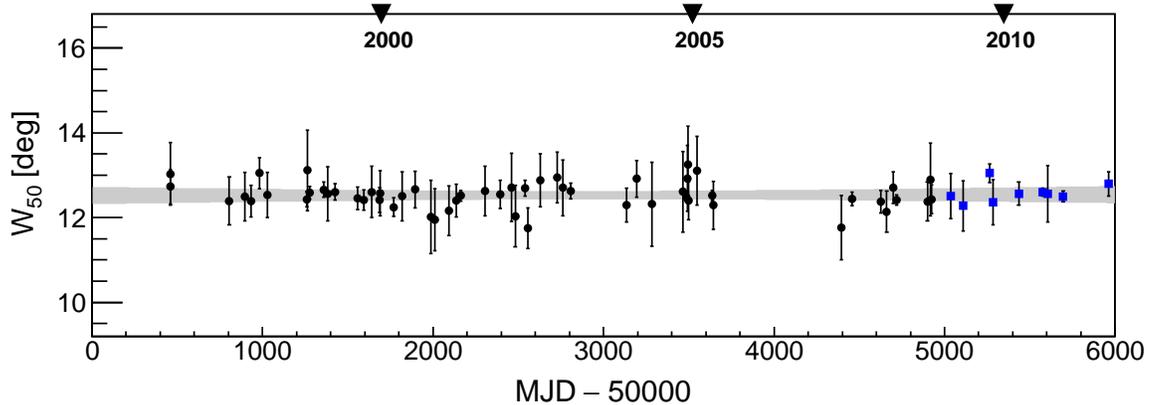}
  \end{center}
\caption{\label{fig:J1744_width} Pulse width at 50\% intensity level
  of \psrj{}, as a function of time. Black circles are observations
  made at 1410\,MHz, while blue squares are observations made at
  1360\,MHz. Errors are rescaled by the square root of the fitting
  $\chi^2_{\rm red}$. The gray region shows the 3-$\sigma$ band of our
  linear fitting. The years of observations are indicated at the top
  of the figure.}
\end{figure}

\psrj{} was discovered in 1997 through the Parkes 436\,MHz survey of
the southern sky~\cite{bjb+97}. It has a spin period of 4.07\,ms, and
later as a target in the PTA projects~\cite{haa+10}, it is being
observed frequently. \psrj{} has a sharp pulse with a $W_{50}\sim
12.5^\circ$ at 1410\,MHz, see figure~\ref{fig:J1744_pulse}. Because of
a long observational span in time and continuous observations since
its discovery, it is also a good laboratory to test the local Lorentz
invariance of gravity.

We used 65 observations spanning about 15 years obtained with the
100-m Effelsberg radio telescope.  In order to measure the pulse width
accurately, we again try to describe the profile by an analytic
function.  The pulse profile of \psrj{} is different from that of
\psrb{}, where we used fits of parabolas to parts of the pulse to
determine the width. The same method cannot be applied here, and
different functions need to be used.  Despite the apparent simplicity
of the pulse, a good fit to a high S/N pulse profile is not trivial,
see e.g. figure 4 in~\cite{mhb+12} where seven Gaussian components
were used for the whole profile at 1400\,MHz observational frequency
(five Gaussian components for the main-pulse). In this work we used
three components to fit the main-pulse of \psrj{}. These components
have close centre values, so using three Gauss functions would often
not give a stable fitting result, unless we a priori fix the means of
them. Therefore we used three components with different shapes (one
Gauss function and two Landau functions with opposite orientations) to
break the degeneracy and achieve a stable fitting. A typical fitting
is shown in the inset of figure~\ref{fig:J1744_pulse}.  For each
observation, we generated $10^4$ realizations of the profile according
to measurement and measurement uncertainty. Then for each profile, we
fitted the three components analytically.  $W_{50}$ is obtained from
the analytical sum of these components. Hence for one observation, we
have a distribution of pulse width with ten thousand entries. A width
with an uncertainty is drawn from this distribution. Finally the
uncertainty is rescaled by the square root of the fitting $\chi^2_{\rm
  red}$.  The result is plotted in figure~\ref{fig:J1744_width} as a
function of time.  No clear evolution over time is seen from the pulse
width. Because of fewer observations and lower S/N of the profile data
on average of \psrj{}, the uncertainties in $W_{50}$ are in general
larger than that of \psrb{}.

\begin{figure}
  \begin{center}
    \includegraphics[width=18cm]{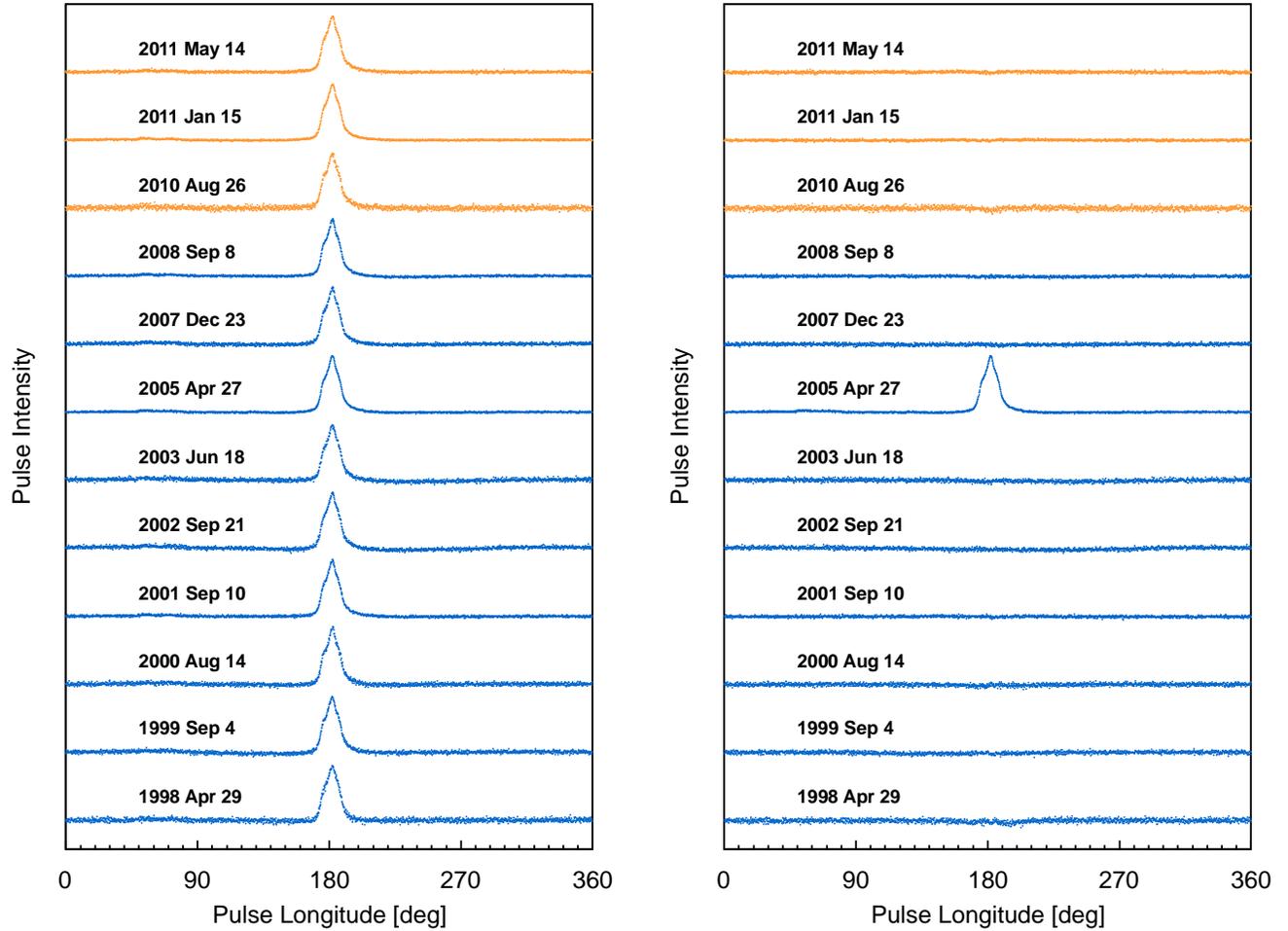}
  \end{center}
\caption{Same as figure~\ref{fig:B1937_diff}, for \psrj{}. For the
  right panel, the pulse profile taken on April 27, 2005 is used as
  the reference profile for subtraction. \label{fig:J1744_diff}}
\end{figure}

Twelve high S/N profiles of \psrj{} are present in
figure~\ref{fig:J1744_diff} in a similar way as \psrb{} in
figure~\ref{fig:B1937_diff}. For this pulsar, we see no evolution
between two frequencies. We performed the same check as with \psrb{}
using {\tt Asterix} data of \psrj{}, and no evolution of the profile
width between two frequencies is found. Hence, we only used one
reference profile for subtraction. From the residuals in the right
panel, one finds no change in pulse profile over $\sim15$\,years.

The same Monte Carlo and fitting analysis applied to \psrb{} (see
section~\ref{sec:psrb}) was implemented for \psrj{}. For this pulsar,
we used the linear function (\ref{eq:linearfitting}) without the need
of a jump in pulse width. The fitting results from $10^6$ simulations
are tabulated in table~\ref{tab:psr}. As was the case for \psrb{}, the
measurement of $(\rmd{W}_{50}/\rmd t)$ shows no significant change in
$W_{50}$ over time for \psrj{}. $F$-test~\cite{ptvf92} gives a
$p$-value of 0.44 for the necessity of a non-zero $(\rmd{W}_{50}/\rmd
t$).

\section{A new $\hatwo$ limit from solitary pulsars}
\label{sec:a2}

From the pulse-profile analysis of \psrs{}, tight limits on the change
of pulse width have been set (see table~\ref{tab:psr}).  We have also
examined the profiles of \psrs{} from early literature and no change
was found. Because of low quanlity of early data and the use of
different backends, they have not been included in the calculations.

By combining (\ref{eq:omegaprec}), (\ref{eq:lambdadot})
and (\ref{eq:w50dot}), we have
\begin{equation}\label{eq:a2}
  \hatwo = -2\,{\cal A} \left[ \frac{2\pi}{P}
    \left(\frac{w}{c}\right)^2 \cos \psi \cos\vartheta \right]^{-1}
  \frac{\rmd W}{\rmd t} \,.
\end{equation}
The limits on width changes for \psrs{} were given in
section~\ref{sec:pulse}.  The others will be discussed in the
following.

${\cal A}$ is defined in (\ref{eq:w50dot}) and includes information of
the pulse profile, the spin orientation and the emission
property. Pulse width is obtained from profile analysis, while
$\lambda$ (or equivalently the observer angle $\zeta \equiv 180^\circ
- \lambda$) and $\alpha$ can be obtained from lightcurve analysis by
combining radio and $\gamma$-ray observations. For both \psrs{},
$\gamma$-ray observations from Fermi Large Area Telescope (LAT) are
available~\cite{gjv+12,joh12}. The results from modeling of the radio
and $\gamma$-ray emission profiles are quoted in table~\ref{tab:psr}
for \psrb{}~\cite{gjv+12} and \psrj{}~\cite{joh12}.

To obtain the quantities inside the square brackets of (\ref{eq:a2}),
besides the well measured spin period (see table~\ref{tab:psr}), we
need to know the pulsar's velocity with respect to the preferred
frame, and the pulsar's spin orientation with respect to it.

First, a preferred frame must be specified. The most natural frame is
the one where the CMB radiation is isotropic. The CMB frame is used as
the preferred frame in most literature (see
however~\cite{wk07,skmb08,sw12} for other local frames), and the
constraints of $\ai$ (and $\hai$) quoted in section~\ref{sec:pfe} all
refer to this frame. We will also use the CMB frame as the preferred
frame in the following calculations. This choice basically assumes
that the preferred frame is determined by the global distribution of
matter in the Universe, and that the fields of the gravitational
interaction, which cause the PFEs, are long range, at least comparable
to the Hubble radius. The generalization to other frames is
straightforward.

From Wilkinson Microwave Anisotropy Probe (WMAP) observations, our
Solar system barycenter (SSB) has a peculiar velocity with respect to
the CMB frame, $|\mathbf{w}_{\mathrm{SSB}}| = 369.0 \pm
0.9~\textrm{km\,s}^{-1}$, in the direction of Galactic longitude and
latitude $(l,b) = (263.99^\circ \pm 0.14^\circ, 48.26^\circ \pm
0.03^\circ)$~\cite{hwh+09,jbd+11}.  The pulsar velocity with respect
to the CMB frame is simply $\mathbf{w} = \mathbf{v}_{\mathrm{PSR-SSB}}
+ \mathbf{w}_{\mathrm{SSB}}$, where $\mathbf{v}_{\mathrm{PSR-SSB}}$ is
the 3D motion of the pulsar with respect to SSB.  The 2D projected
movement on the sky plane can be obtained from proper motion and
parallax measurements from timing experiments~\cite{vbc+09} (see
table~\ref{tab:psr}).  The parallax of \psrb{} is not well measured,
so the distance estimated from it is not accurate. Different Galactic
electron density models~\cite{tc93,cl02,sch12} infer a distance in the
range of $3.6$--$4.8$\,kpc (D.~Schnitzeler, private communication),
coarsely consistent with the distance derived from the parallax.
Fortunately, because of the small angular proper motion of \psrb{}
($\mu_{\rm T} \equiv \sqrt{\mu_\alpha^2+\mu_\delta^2}\simeq0.42{\rm
  \,mas\,yr}^{-1}$; see table~\ref{tab:psr}), the error of the 2D
velocity is less than $10{\rm \,km\,s}^{-1}$ even if we underestimate
or overestimate the distance by a few kpc.  The radial velocity $v_r
\equiv \hat{\bf K} \cdot{\bf v}_{\rm PSR-SSB}$ of solitary pulsars in
general is not measurable from pulsar timing experiments. However, we
can see in the following that it only has slight effects on the
test. The radial velocity enters in (\ref{eq:a2}) through ${\bf w}$,
in the form of $({\bf w}\cdot \hat{\bf s})$\footnote{It has no
  contribution to $({\bf w}\cdot \hat{\bf e})$ because by definition
  $\hat{\bf e}$ is in the sky plane.}. From Fermi LAT $\gamma$-ray
observations, we can see that the spins of \psrs{} both lie close to
the sky plane ($\zeta\sim80^\circ$; see table~\ref{tab:psr}), so the
unknown radial velocity only has a marginal effect in $({\bf w}\cdot
\hat{\bf s})$.  By assuming that the solitary pulsars are
gravitationally bound in the Galaxy~\cite{kg89}, we find that the
reasonable ranges of the radial velocities are $-600\,{\rm km\,s}^{-1}
\lesssim v_r \lesssim 200\,{\rm km\,s}^{-1}$ for \psrb{} and
$-400\,{\rm km\,s}^{-1} \lesssim v_r \lesssim 250\,{\rm km\,s}^{-1}$
for \psrj{}, respectively.  We use these ranges to test the dependence
of our $\hatwo$ limits on the radial velocity later on, and the
results only show a weak dependence that alters the limits by
$\sim15\%$ at most. Even if we assume some unphysical radial velocity
$|v_r| \gtrsim 1000\,{\rm km\,s}^{-1}$, the limits are altered by
$\sim40\%$ at most. In the case of extremely large radial velocities
$|v_r| \gtrsim 1500\,{\rm km\,s}^{-1}$, the $\hatwo$ limits get better
with increasing $|v_r|$.

For the pulsar spin orientation, as mentioned before, $\zeta$ can be
inferred from the combination of radio and Fermi LAT observations.
The remaining unknown is the azimuthal angle $\eta$ of the pulsar spin
$\hat{\bf s}$ around the line of sight (see
figure~\ref{fig:geometry}), which is not an observable from pulsar
observations for \psrs{} and will be treated as a random variable.

\begin{figure}
\begin{center}
\includegraphics[width=13cm]{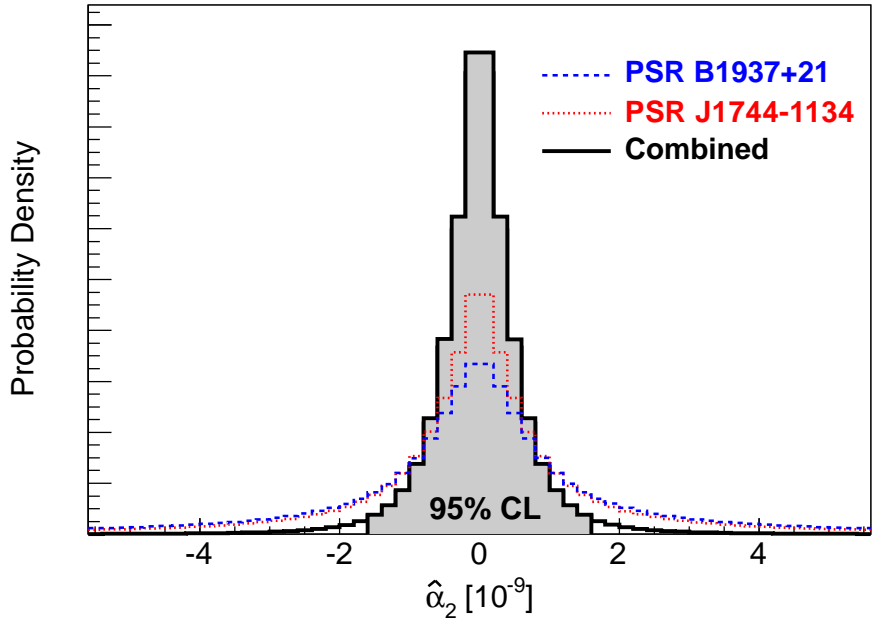}
\end{center}
  \caption{\label{fig:pdf} Probability density functions of $\hatwo$
    from \psrb{} (blue dashed histogram), \psrj{} (red dotted
    histogram), and their combination (black solid histogram). At 95\%
    confidence level, $|\hatwo|$ is constrained to be less than
    $1.6\times10^{-9}$ from the combined probability distribution.}
\end{figure}

We set up Monte Carlo simulations to account for measurement errors
and the unknown $\eta$ and the unknown radial velocity.  In our
simulation, we assume that the radial velocity follows a Gaussian
distribution with a zero mean and a $100\,{\rm km}\,{\rm s}^{-1}$
spread, and $\eta$ is treated as a random variable uniformly
distributed in $(0^\circ,360^\circ)$, in the same way as that
of~\cite{ds91,de92}. As mentioned before, we also set up various Monte
Carlo simulations for different radial velocities, where only a weak
dependence on the choice of the radial velocity is found, at most
altering our results by $\sim15\%$ under the assumption that the
solitary pulsars are bound in the gravitational potential of the
Galaxy.  Because of the unknown $\eta$, our final result is a
probabilistic constraint, the same as the strong equivalence principle
test in~\cite{ds91} and the $\haone$ test in~\cite{de92}. It is the
main limitation of these tests (see~\cite{sw12} for a robust $\haone$
test where such a probabilistic assumption was overcome).  Through
$10^8$ simulations, we got the probability density functions (PDFs) of
$\hatwo$ from \psrs{} according to (\ref{eq:a2}). They are plotted in
figure~\ref{fig:pdf} as a blue dashed histogram and a red dotted
histogram, respectively. From these PDFs, we obtain
\begin{eqnarray}
  \mbox{\psrb{}:} &\quad& 
  |\hatwo| < 2.5 \times 10^{-8}\,, \quad \mbox{(95\% CL)},\label{eq:a2psrb} \\
  \mbox{\psrj{}:} &\quad&
  |\hatwo| < 1.5 \times 10^{-8}\,, \quad \mbox{(95\% CL)}.\label{eq:a2psrj}
\end{eqnarray}
They are already much better than the limit (\ref{eq:a2solar}) from
the Solar system. For these limits, \psrb{} benefits from a smaller
spin period and a tighter constraint on $\rmd{W}_{50}/\rmd t$ (see
table~\ref{tab:psr}), however \psrj{} benefits from a more favorable
emission geometry (a smaller ${\cal A}$). In total, \psrj{} gives a
slightly better limit than \psrb{}. The analysis for \psrb{} is based
on the main-pulse ({\tt MP1} in figure
\ref{fig:B1937_pulse}). Likewise, one could use the interpulse ({\tt
  IP} in figure \ref{fig:B1937_pulse}) to constrain a precession of
\psrb{}, which leads to a similar, even slightly more constraining
limit because of a smaller ${\cal A}$.  We therefore stay with the
more conservative value derived from the main-pulse.  Similarly, even
though the results at $\gamma$-ray frequencies convincingly rule out
such an interpretation, one may consider the main- and interpulse as
the result of a single very wide cone. In that case, the
interpretation of the change in width as described in
(\ref{eq:w50dot}) will need to be recasted in terms of the IP-MP
separation ({\tt SEP0} in figures~\ref{fig:B1937_pulse} and
\ref{fig:B1937_width}). Such an interpretion would give similar
limits.

We can immediately see that the above two numbers are located far
outside the $\hatwo$ range plotted in figure~\ref{fig:pdf}. This is
due to the fact that these PDFs have very long tails (compared with
the normal distribution). The reason for the long tail was analyzed
explicitly in~\cite{sw12} for a similar $\hatwo$ test from binary
dynamics. They are due to unfavorable geometrical configurations where
$\cos\psi \simeq 0$ and/or $\cos\vartheta\simeq 0$. From
(\ref{eq:a2}), we can see that $\hatwo$ is unconstrained at these
configurations. Fortunately, as in~\cite{sw12} one can take advantage
of the probabilistic consideration by using more than one system to
suppress the long tails. The probability that {\it both} pulsars are
at their unfavorable orientation is small.  For this reason we use
more than one solitary pulsar. By assuming that \psrs{} are
independent and that they have approximately the same $\hatwo$ value,
we got a combined PDF for $\hatwo$.  It is shown in
figure~\ref{fig:pdf} as a solid black histogram. The long tail is
highly suppressed as one expects. From the combined PDF, we get
\begin{equation}\label{eq:a2psr}
  |\hatwo| < 1.6 \times 10^{-9} \,, \quad \mbox{(95\% CL)}
\end{equation}
which is significantly better than that of (\ref{eq:a2solar}) from the
Solar system~\cite{nor87}, and more than four orders of magnitude
better than the limit (\ref{eq:a2llr}) from LLR~\cite{mwt08}.

\begin{figure}
\begin{center}
\includegraphics[width=13cm]{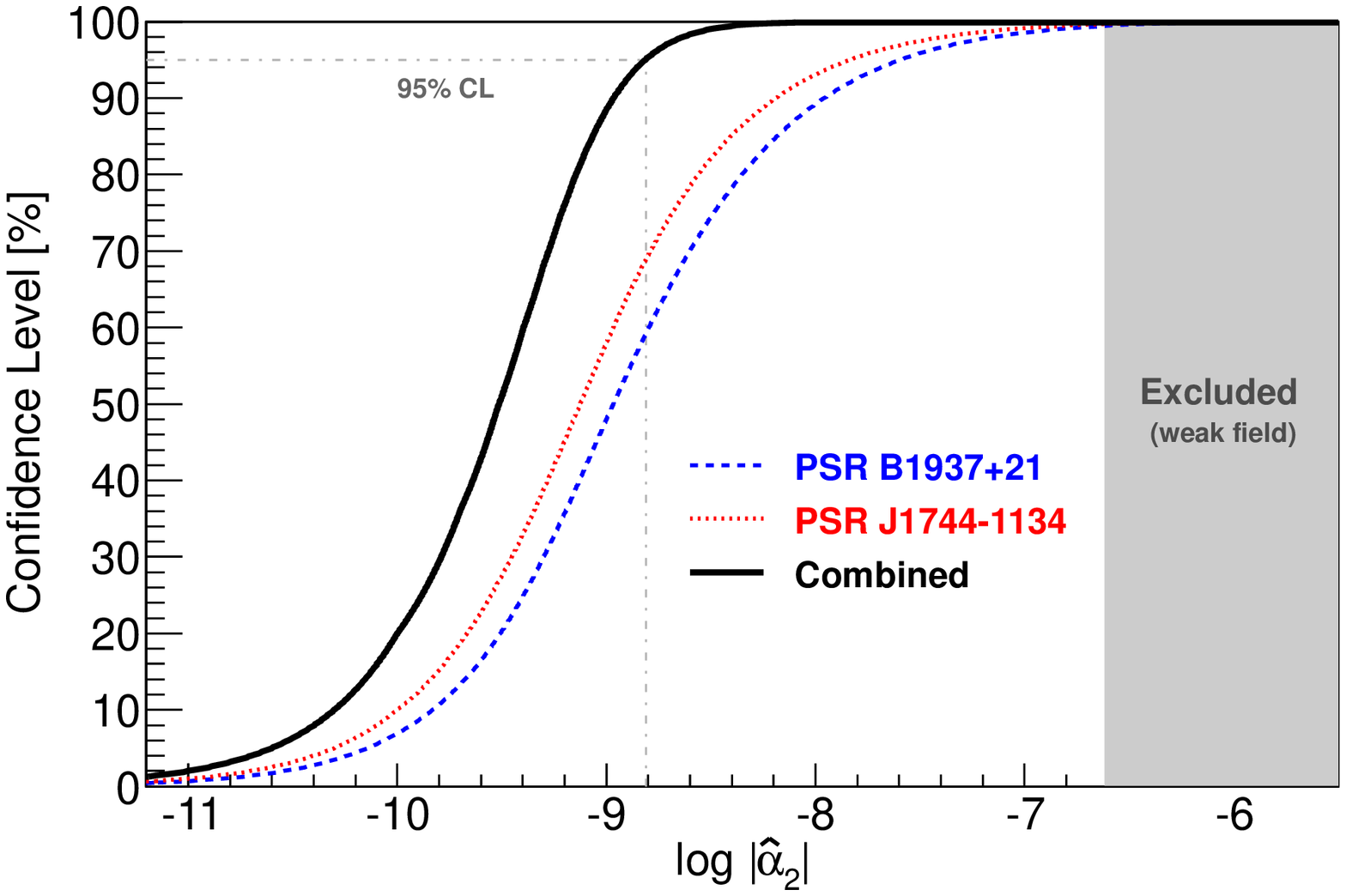}
\end{center} 
  \caption{\label{fig:logpdf} Comparisons of our limits on $|\hatwo|$
    with that from the Solar system~\cite{nor87}. The curves show the
    confidence levels to statistically reject such an $|\hatwo|$
    according to the measurements from \psrb{} (blue dashed line),
    \psrj{} (red dotted line), and their combination (black solid
    line). The solar system constraint (\ref{eq:a2solar}) on the
    weak-field $\atwo$~\cite{nor87} is illustrated as the gray region,
    and the combined limit (\ref{eq:a2psr}) at 95\% confidence level
    is also indicated.}
\end{figure}

To compare our results with the limit (\ref{eq:a2solar}) graphically,
a logarithmic scale is needed. We plot in figure~\ref{fig:logpdf} the
confidence levels to exclude a specific $\hatwo$ value versus ${\rm
  log}\,|\hatwo|$, from \psrs{} and their combination. The limit
(\ref{eq:a2solar}) is plotted as an exclusion region in gray. The
improvement of the limit by orders of magnitude is obvious.

\section{Discussions}
\label{sec:con}

Strictly speaking, the comparison between (\ref{eq:a2solar}) and
(\ref{eq:a2psr}) is only phenomenological. $\atwo$ and $\hatwo$ probe
different aspects of the local Lorentz symmetry of gravity, namely
weak fields and strong fields. It was explicitly shown that in the
strong fields, one can have a different PPN parameter in the
scalar-tensor theories through a mechanism called ``scalarization''
similar to the well known phenomenon ``phase transition''~\cite{de93}.
As an example, in the scalar-tensor gravity the PPN parameter $\gamma$
generalizes to
\begin{equation}
  \hat\gamma \equiv \gamma_{AB} = 
  1 - \frac{2\alpha_A\alpha_B}{1 + \alpha_A\alpha_B} \,,
\end{equation}
for a binary pulsar system, where $\alpha_A$ and $\alpha_B$ are the
effective scalar coupling constants of the pulsar and its companion,
respectively~\cite{de92a}.  The weak-field PPN parameter $\gamma$ is
recovered for $\alpha_A = \alpha_B = \alpha_0$. In GR one has
$\hat\gamma = \gamma = 1$. Similarly, we may expect that
$\hat\alpha_2$ deviates from its weak-field PPN correspondent
$\alpha_2$ due to strong-field contributions. The strong-field
$\hatwo$ in the Einstein-\AE{}ther theory can be found
in~\cite{fos07}.  In the absence of non-perturbative effects, one can
expand $\hatwo$ in the compactness ${\cal C}$ of the
body~\cite{de96}. In our case, we would write an expansion like,
\begin{equation}\label{eq:expanse}
  \hat\alpha_2 = \alpha_2 + {\cal K}_1 {\cal C} + {\cal K}_2 {\cal
    C}^2 + \dots \,,
\end{equation}
where ${\cal K}_i$ are coefficients characterizing deviations from GR,
and ${\cal C} \simeq GM/Rc^2$ for a body with mass $M$ and radius
$R$. The compactnesses for the Earth and the Sun are roughly ${\cal
  C}_\oplus \sim 10^{-9}$ and ${\cal C}_\odot \sim 10^{-6}$,
respectively, which suppress ${\cal K}_i$-related ($i=1,2,\cdots$)
physical effects dramatically. In contrast, neutron stars have ${\cal
  C}_{\rm NS} \sim 0.2$, which is one of the reasons why pulsars are
ideal probes for gravity effects associated with strong gravitational
fields.  From (\ref{eq:expanse}) one can see that, the $\hatwo$ limit
from pulsars is $\sim 10^{5}$ times more sensitive to the ${\cal K}_1$
parameter and $\sim 10^{10}$ times more sensitive to the ${\cal K}_2$
parameter than that of the Solar system test~(\ref{eq:a2solar}). It is
even more sensitive compared with the constraint~(\ref{eq:a2llr}) from
the ranging experiment of the Sun-Earth-Moon dynamics.  This supports
the importance of the strong-field $\hat\alpha_2$ limit
(\ref{eq:a2psr}).

It is worth mentioning that, when discussing the constraints on
strong-field parameters of alternative gravity theories, one should be
aware of a potential compactness-dependent nature of these parameters,
especially when combining different systems. Our $\hatwo$ test
(\ref{eq:a2psr}) assumes that $\hatwo$ is approximately the same for
\psrs{}.  However, in the presence of phenomena related to some
critical mass, like the spontaneous scalarization discovered in the
scalar-tensor theory~\cite{de93}, even a small difference in masses
does not allow such an assumption (see~\cite{de96b,fwe+12,afw+13} for
constraints on such critical phenomena). Therefore, the comparison
between (\ref{eq:a2solar}), (\ref{eq:a2llr}), (\ref{eq:a2bnrypsr}) and
(\ref{eq:a2psr}) is only phenomenological. More strictly, they measure
different aspects of gravity under different circumstances, such as
the gravitational environments of the Sun, the Sun-Earth-Moon orbital
dynamics, the pulsar-white dwarf orbital dynamics, and the solitary
pulsars.

The main result of the paper (\ref{eq:a2psr}) is phenomenological in
the PPN framework, nevertheless, it is relevant to some alternative
gravity theories with local Lorentz invariance violation, like the
Ho{\v r}ava-Lifshitz gravity~\cite{hor09,bps11}, the
Einstein-\AE{}ther theory~\cite{jm01,fj06} and the TeVeS
theory~\cite{bek04,sag09}. A detailed comparison with these
alternative gravity theories has to account for possible strong-field
dependencies.  Such an analysis is beyond the scope of this paper.

As one can see from figures~\ref{fig:pdf} and \ref{fig:logpdf}, the
improvement of the combined limit over that from a single pulsar is
significant. The combined limit (\ref{eq:a2psr}) is ten times better
than the one solely obtained from \psrb{} or \psrj{}.  This reveals
the probabilistic consideration inherent as well as the directional
dependence of PFEs. When pulsars located differently and moving with
different velocities in different directions are used, the constraint
on PFEs is much stronger than solely from one object. It is also
demonstrated in~\cite{wk07} for binary pulsars under a similar
concept, called ``PFE antenna array''. In the $\hatwo$ test, we
include two solitary pulsars with highest figure of merit (see
below). Through our simulations, we found that by including more
pulsars with lower figure of merit, the improvement is not
significant. However, two pulsars are the minimum requirement to
suppress the long tails discussed earlier.

The figure of merit of the $\hatwo$ test proposed here can be
extracted from (\ref{eq:a2}). In general, it depends on the upper
limit on the change of pulse width, the spin period, the ``absolute''
velocity ${\bf w}$, the pulsar's spin orientation with respect to
${\bf w}$ and the line of sight, and also some emission properties
encoded in ${\cal A}$. After dropping complicated dependence, one can
see that the power of the test is roughly proportional to $\left[ P
  \left({{\rmd W}/{\rmd t}}\right)^{\rm upper} \right]^{-1}$, where
$\left(\rmd{W}/\rmd t\right)^{\rm upper}$ is the upper limit of the
change in the pulse width and $P$ is the pulsar spin period. Hence one
can see that the solitary pulsars with short spin period and smaller
$\left(\rmd{W}/\rmd t\right)^{\rm upper}$ are more useful in setting a
tight constraint of $\hatwo$\footnote{However, in general, all
  dependencies in (\ref{eq:a2}) contribute; for example, although
  \psrb{} has a higher $\left[ P \left({{\rmd W}/{\rmd t}}\right)^{\rm
      upper} \right]^{-1}$ compared with \psrj{}, it gets a slightly
  worse constraint on $\hatwo$ than \psrj{} because of a significantly
  larger ${\cal A}$.}. The quantity $\left(\rmd{W}/\rmd t\right)^{\rm
  upper}$ can depend on different factors, like the luminosity of the
emission, and the pulse width.  If we conservatively assume no
improvement in the observational technologies, it scales roughly as
$T_{\rm obs}^{-3/2}$, where $T_{\rm obs}$ is the observational time
span. Hence finding more pulsars with short spin period, and
continuous observations on known MSPs both help in improving the
$\hatwo$ limit.  In the era of new telescopes, like the
Five-hundred-meter Aperture Spherical Telescope (FAST)~\cite{nlj+11}
and the Square Kilometre Array (SKA)~\cite{sks+09}, with more
dedicated technologies, more pulsars are to be found for sure and data
with better quality are guaranteed.  On the other hand, many stable
MSPs are also used in the PTA projects~\cite{fhb+10,hlj+11,haa+10} and
being observed continuously (like \psrs{}), hence the $\hatwo$ test
proposed here can be achieved as a byproduct from other science
programs, and is expected to improve continuously.

In summary, we proposed to use the non-detection of spin precession of
solitary pulsars from pulse profile analysis to constrain the
strong-field PPN parameter $\hatwo$. Two solitary pulsars, \psrs{},
are used to get a combined limit of $|\hatwo| < 1.6 \times 10^{-9}$ at
95\% confidence level (see (\ref{eq:a2psr})), which is significantly
better than the limit (\ref{eq:a2solar}) obtained from the Solar
system~\cite{nor87}. Moreover, the $\hatwo$ test with solitary pulsars
is based on regular observations over the whole time span, excluding
for instance a 360$^\circ$ precession between the starting and the end
points.  In contrast to the Solar limit, our test will continuously
improve the limit from finding new pulsars as well as long-term
regular observations on known pulsars.

\section*{Acknowledgments}

We are grateful to Kosmas Lazaridis and Oliver L\"ohmer for helping
with observations at the 100-m Effelsberg radio telescope.  We thank
Ewan Barr, Paulo Freire, Lucas Guillemot, Patrick Lazarus, K.J. Lee,
Dominic Schnitzeler, and Joris Verbiest for discussions. We also thank
Patrick Lazarus for helping with the reduction of {\tt Asterix} data
and Paulo Freire for reading the manuscript. LS is supported by China
Scholarship Council (CSC). RNC is a member of the International Max
Planck Research School (IMPRS) for Astronomy and Astrophysics at the
Universities of Bonn and Cologne. This research has made use of NASA's
Astrophysics Data System.



\section*{References}

\begin{thebibliography}{10}

\bibitem{afw+13}
J.~Antoniadis, P.~C.~C.~Freire, N.~Wex, \etal.
\newblock {A Massive Pulsar in a Compact Relativistic Binary}.
\newblock {\em Science}, 340:448, 2013.


\bibitem{bdz+97}
D.~C.~Backer, M.~R.~Dexter, A.~Zepka, \etal.
\newblock {A Programmable 36-MHz Digital Filter Bank for Radio Science}.
\newblock {\em Publications of the Astronomical Society of the
  Pacific}, 109:61--68, 1997.

\bibitem{bkh+82}
D.~C. {Backer}, S.~R. {Kulkarni}, C.~{Heiles}, 
\newblock {A millisecond pulsar}.
\newblock {\em \nat}, 300:615--618, 1982.

\bibitem{bai88}
M.~{Bailes}.
\newblock {Geodetic precession in binary pulsars}.
\newblock {\em \aap}, 202:109--112, 1988.

\bibitem{bjb+97}
M.~{Bailes}, S.~{Johnston}, J.~F. {Bell}, 
  \etal.
\newblock {Discovery of Four Isolated Millisecond Pulsars}.
\newblock {\em \apj}, 481:386, 1997.

\bibitem{bk06}
Q.~G. {Bailey} and V.~A. {Kosteleck{\'y}}.
\newblock {Signals for Lorentz violation in post-Newtonian gravity}.
\newblock {\em \prd}, 74:045001, 2006.

\bibitem{bek04}
J.~D.~Bekenstein.
\newblock {Relativistic gravitation theory for the modified Newtonian
  dynamics paradigm}.
\newblock {\em \prd}, 70:083509, 2004. [Erratum: {\em \prd}, 71:069901, 2005]

\bibitem{bcc+04}
N.~D.~R.~Bhat, J.~M.~Cordes, F.~Camilo, D.~J.~Nice, and D.~R.~Lorimer.
\newblock {Multifrequency Observations of Radio Pulse Broadening and
  Constraints on Interstellar Electron Density Microstructure}.
\newblock {\em \apj}, 605:759--783, 2004.

\bibitem{bps11}
D.~{Blas}, O.~{Pujol{\`a}s}, and S.~{Sibiryakov}.
\newblock {Models of non-relativistic quantum gravity: the good, the bad and
  the healthy}.
\newblock {\em Journal of High Energy Physics}, 4:18, 2011.

\bibitem{cl02}
J.~M.~Cordes and T.~J.~W.~Lazio.
\newblock {NE2001.I. A New Model for the Galactic Distribution of Free
Electrons and its Fluctuations}.
arXiv:astro-ph/0207156, 2002.

\bibitem{cwb90}
J.~M. {Cordes}, I.~{Wasserman}, and M.~{Blaskiewicz}.
\newblock {Polarization of the binary radio pulsar 1913+16 --- constraints on
  geodetic precession}.
\newblock {\em \apj}, 349:546--552, 1990.

\bibitem{de92}
T.~{Damour} and G.~{Esposito-Far{\`e}se}.
\newblock {Testing local Lorentz invariance of gravity with binary-pulsar
  data}.
\newblock {\em \prd}, 46:4128--4132, 1992.

\bibitem{de92a}
T.~{Damour} and G.~{Esposito-Far{\`e}se}.
\newblock {Tensor-multi-scalar theories of gravitation}.
\newblock {\em Classical and Quantum Gravity}, 9:2093--2176, 1992.

\bibitem{de93}
T.~{Damour} and G.~{Esposito-Far{\`e}se}.
\newblock {Nonperturbative strong-field effects in tensor-scalar theories of
  gravitation}.
\newblock {\em Physical Review Letters}, 70:2220--2223, 1993.

\bibitem{de96}
T.~{Damour} and G.~{Esposito-Far{\`e}se}.
\newblock {Testing gravity to second post-Newtonian order: A field-theory
  approach}.
\newblock {\em \prd}, 53:5541--5578, 1996.

\bibitem{de96b}
T.~{Damour} and G.~{Esposito-Far{\`e}se}.
\newblock {Tensor-scalar gravity and binary-pulsar experiments}.
\newblock {\em \prd}, 54:1474--1491, 1996.

\bibitem{ds91}
T.~{Damour} and G.~{Sch\"afer}.
\newblock {New tests of the strong equivalence principle using binary-pulsar
  data}.
\newblock {\em Physical Review Letters}, 66:2549--2552, 1991.

\bibitem{dr75}
N.~D.~H.~Dass and V.~Radhakrishnan.
\newblock {The new binary pulsar and the observation of
      gravitational spin precession}.
\newblock {\em Astrophysical Letters}, 16:135--139, 1975.

\bibitem{fos07}
B.~Z. {Foster}.
\newblock {Strong field effects on binary systems in Einstein-aether theory}.
\newblock {\em \prd}, 76:084033, 2007.

\bibitem{fhb+10}
R.~D.~Ferdman, R.~van~Haasteren, C.~G.~Bassa, \etal.
\newblock {The European Pulsar Timing Array: current efforts and a
  LEAP toward the future}.
\newblock {\em Classical and Quantum Gravity}, 27:084014, 2010.

\bibitem{fj06}
B.~Z. {Foster} and T.~{Jacobson}.
\newblock {Post-Newtonian parameters and constraints on Einstein-aether
  theory}.
\newblock {\em \prd}, 73:064015, 2006.

\bibitem{fwe+12}
P.~C.~C.~Freire, N.~Wex, G.~Esposito-Far\`{e}se, \etal.
\newblock {The relativistic pulsar-white dwarf binary PSR J1738+0333 -
II. The most stringent test of scalar-tensor gravity}.
\newblock {\em \mnras}, 423:3328, 2012.

\bibitem{ggr84}
J.~{Gil}, P.~{Gronkowski}, and W.~{Rudnicki}.
\newblock {Geometry of the emission region of PSR 0950+08}.
\newblock {\em \aap}, 132:312--316, 1984.

\bibitem{gjv+12}
L.~{Guillemot}, T.~J. {Johnson}, C.~{Venter}, 
  \etal.
\newblock {Pulsed Gamma Rays from the Original Millisecond and Black Widow
  Pulsars: A Case for Caustic Radio Emission?}
\newblock {\em \apj}, 744:33, 2012.

\bibitem{hlj+11}
R.~van~Haasteren, Y.~Levin, G.~H.~Janssen, \etal.
\newblock {Placing limits on the stochastic gravitational-wave
  background using European Pulsar Timing Array data}.
\newblock {\em \mnras}, 414:3117, 2011. [Erratum: {\em \mnras},
  425:1597, 2012]

\bibitem{hbp+68}
A.~{Hewish}, S.~J. {Bell}, J.~D.~H. {Pilkington}, P.~F. {Scott}, and R.~A.
  {Collins}.
\newblock {Observation of a Rapidly Pulsating Radio Source}.
\newblock {\em \nat}, 217:709--713, 1968.

\bibitem{hwh+09}
G.~{Hinshaw}, J.~L. {Weiland}, R.~S. {Hill}, \etal.
\newblock {Five-Year Wilkinson Microwave Anisotropy Probe Observations: Data
  Processing, Sky Maps, and Basic Results}.
\newblock {\em \apjs}, 180:225--245, 2009.

\bibitem{haa+10}
G.~{Hobbs}, A.~{Archibald}, Z.~{Arzoumanian}, 
  \etal.
\newblock {The International Pulsar Timing Array project: using pulsars as a
  gravitational wave detector}.
\newblock {\em Classical and Quantum Gravity}, 27:084013, 2010.

\bibitem{hcm+12}
G.~{Hobbs}, W.~{Coles}, R.~N. {Manchester}, 
  \etal.
\newblock {Development of a pulsar-based timescale}.
\newblock {\em \mnras}, 427:2780, 2012.

\bibitem{hor09}
P.~{Ho{\v r}ava}.
\newblock {Quantum gravity at a Lifshitz point}.
\newblock {\em \prd}, 79:084008, 2009.

\bibitem{hsm04}
A.~W.~Hotan, W.~van~Straten, and R.~N.~{Manchester}.
\newblock {PSRCHIVE and PSRFITS: An Open Approach to Radio Pulsar Data
Storage and Analysis}.
\newblock {\em Publications of the Astronomical Society of Australia},
21:302--309, 2004.

\bibitem{jm01}
T.~{Jacobson} and D.~{Mattingly}.
\newblock {Gravity with a dynamical preferred frame}.
\newblock {\em \prd}, 64:024028, 2001.

\bibitem{jbd+11}
N.~{Jarosik}, C.~L. {Bennett}, J.~{Dunkley}, \etal.
\newblock {Seven-year Wilkinson Microwave Anisotropy Probe (WMAP) Observations:
  Sky Maps, Systematic Errors, and Basic Results}.
\newblock {\em \apjs}, 192:14, 2011.

\bibitem{joh12}
T.~J. {Johnson}.
\newblock {Constraints on the Emission Geometries of Gamma-ray Millisecond
  Pulsars Observed with the Fermi Large Area Telescope}.
\newblock {\em PhD thesis}, University of Maryland, 2011. [arXiv:1209.4000]

\bibitem{kg89}
K.~Kuijken and G.~Gilmore.
\newblock {The Mass Distribution in the Galactic Disc -- II.
  Determination of the Surface Mass Density of the Galactic Disc Near
  the Sun}.
\newblock {\em \mnras}, 239:605--649, 1989.

\bibitem{kos04}
V.~A.~Kosteleck\'{y}.
\newblock {Gravity, Lorentz violation, and the standard model}.
\newblock {\em \prd}, 69:105009, 2004.

\bibitem{kra98}
M.~{Kramer}.
\newblock {Determination of the Geometry of the PSR B1913+16 System by Geodetic
  Precession}.
\newblock {\em \apj}, 509:856--860, 1998.

\bibitem{kll+99}
M.~{Kramer}, C.~{Lange}, D.~R. {Lorimer}, 
  \etal.
\newblock {The Characteristics of Millisecond Pulsar Emission. III. From Low to
  High Frequencies}.
\newblock {\em \apj}, 526:957--975, 1999.


\bibitem{ksm+06a}
M.~{Kramer}, I.~H. {Stairs}, R.~N. {Manchester}, \etal.
\newblock {Tests of General Relativity from Timing the Double Pulsar}.
\newblock {\em Science}, 314:97--102, 2006.


\bibitem{lk05}
D.~R. {Lorimer} and M.~{Kramer}.
\newblock {\em {Handbook of Pulsar Astronomy}}. Cambridge University Press.
\newblock 2005.


\bibitem{mhb+12}
R.~N. {Manchester}, G.~{Hobbs}, M.~{Bailes}, \etal.
\newblock {The Parkes Pulsar Timing Array Project}.
\newblock {\em Publications of the Astronomical Society of Australia
}, 30:017, 2013.

\bibitem{mhth05} 
R.~N.~Manchester, G.~B.~Hobbs, A.~Teoh, and M.~Hobbs.
  \newblock {The Australia Telescope National Facility Pulsar
    Catalogue}.  \newblock {\em AJ}, 129:1993--2006, 2005.

\bibitem{mkp+05}
R.~N. {Manchester}, M.~{Kramer}, A.~Possenti, \etal.
\newblock {The Mean Pulse Profile of PSR J0737$-$3039A}.
\newblock {\em \apj}, 621:L49--L52, 2005.

\bibitem{mks+10}
R.~N. {Manchester}, M.~{Kramer}, I.~H. {Stairs}, \etal.
\newblock {Observations and Modeling of Relativistic Spin Precession in PSR
  J1141$-$6545}.
\newblock {\em \apj}, 710:1694--1709, 2010.

\bibitem{mwt08}
J.~{M{\"u}ller}, J.~G. {Williams}, and S.~G. {Turyshev}.
\newblock {Lunar Laser Ranging Contributions to Relativity and Geodesy}.
\newblock In {H.~Dittus, C.~Lammerzahl, \& S.~G.~Turyshev}, editor, {\em
  Lasers, Clocks and Drag-Free Control: Exploration of Relativistic Gravity in
  Space}, volume 349 of {\em Astrophysics and Space Science Library}, page 457,
  2008.

\bibitem{nlj+11}
R.~{Nan}, D.~{Li}, C.~{Jin}, \etal.
\newblock {The Five-Hundred Aperture Spherical Radio Telescope (FAST) Project}.
\newblock {\em International Journal of Modern Physics D}, 20:989--1024, 2011.

\bibitem{nor87}
K.~{Nordtvedt}.
\newblock {Probing gravity to the second post-Newtonian order and to one part
  in $10^7$ using the spin axis of the sun}.
\newblock {\em \apj}, 320:871--874, 1987.

\bibitem{nw72}
K.~{Nordtvedt}, and C.~M. {Will}.
\newblock {Conservation Laws and Preferred Frames in Relativistic Gravity. II.
  Experimental Evidence to Rule Out Preferred-Frame Theories of Gravity}.
\newblock {\em \apj}, 177:775, 1972.

\bibitem{pmk+10}
B. B. P. Perera, M.~A.~McLaughlin, M. Kramer, \etal.
\newblock {The Evolution of PSR J0737$-$3039B and a Model for
  Relativistic Spin Precession}.
\newblock {\em \apj}, 721:1193--1205, 2010.

\bibitem{ptvf92}
W.~H.~Press, S.~A.~Teukolsky, W.~T.~Vetterling, and
B.~P.~Flannery.
\newblock{\it Numerical recipes in FORTRAN.}
Cambridge University Press, 1992.

\bibitem{rdb+06}
R.~Ramachandran, P.~Demorest, D.~C.~Backer, I.~Cognard, and A.~Lommen.
\newblock {Interstellar Plasma Weather Effects in Long-Term
  Multifrequency Timing of Pulsar B1937+21}.
\newblock {\em \apj}, 645:303, 2006.

\bibitem{sag09}
E.~{Sagi}.
\newblock {Preferred frame parameters in the tensor-vector-scalar theory of
  gravity and its generalization}.
\newblock {\em \prd}, 80:044032, 2009.

\bibitem{sch12}
D.~H.~F.~M.~Schnitzeler.
\newblock {Modelling the Galactic distribution of free electrons}.
\newblock {\em MNRAS}, 427:664--678, 2012.

\bibitem{sw12}
L.~{Shao} and N.~{Wex}.
\newblock {New tests of local Lorentz invariance of gravity with
  small-eccentricity binary pulsars}.
\newblock {\em Classical and Quantum Gravity}, 29:215018, 2012.

\bibitem{sks+09}
R.~{Smits}, M.~{Kramer}, B.~{Stappers}, \etal.
\newblock {Pulsar searches and timing with the square kilometre array}.
\newblock {\em \aap}, 493:1161--1170, 2009.

\bibitem{skmb08}
M.~{Soffel}, S.~{Klioner}, J.~{M\"uller} and L.~{Biskupek}.
\newblock {Gravitomagnetism
and lunar laser ranging}.
\newblock {\em \prd}, 78:024033, 2008.

\bibitem{sta04}
I.~H. {Stairs}, S.~E.~Thorsett, and Z.~Arzoumanian.
\newblock {Measurement of Gravitational Spin-Orbit Coupling in a
  Binary-Pulsar System}.
\newblock {\em Physical Review Letters}, 93:141101, 2004.

\bibitem{sfl+05}
I.~H. {Stairs}, A.~J. {Faulkner}, A.~G. {Lyne}, \etal.
\newblock {Discovery of Three Wide-Orbit Binary Pulsars: Implications for
  Binary Evolution and Equivalence Principles}.
\newblock {\em \apj}, 632:1060--1068, 2005.

\bibitem{tc93} 
J.~H.~Taylor and J.~M.~Cordes.  
\newblock {Pulsar
  distances and the galactic distribution of free electrons}.
\newblock {\em \apj}, 411:674--684, 1993.

\bibitem{tfm79}
J.~H.~Taylor, L.~A.~Fowler, and P.~M.~McCulloch.
\newblock {Measurements of general relativistic effects in the binary
  pulsar PSR 1913+16}.
\newblock {\em \nat}, 277:437--440, 1979.

\bibitem{tw82}
J.~H. {Taylor} and J.~M. {Weisberg}.
\newblock {A new test of general relativity --- Gravitational radiation and the
  binary pulsar PSR 1913+16}.
\newblock {\em \apj}, 253:908--920, 1982.

\bibitem{vbc+09}
J.~P.~W. {Verbiest}, M.~{Bailes}, W.~A. {Coles}, \etal.
\newblock {Timing stability of millisecond pulsars and prospects for
  gravitational-wave detection}.
\newblock {\em \mnras}, 400:951--968, 2009.

\bibitem{vlm10}
J.~P.~W. {Verbiest}, D.~R.~Lorimer, and M.~A.~McLaughlin.
\newblock {Lutz--Kelker bias in pulsar parallax measurements}.
\newblock {\em \mnras}, 405:564--572, 2010.


\bibitem{wrt89}
J.~M. {Weisberg}, R.~W. {Romani}, and J.~H. {Taylor}.
\newblock {Evidence for geodetic spin precession in the binary pulsar 1913+16}.
\newblock {\em \apj}, 347:1030--1033, 1989.

\bibitem{wk07}
N.~{Wex} and M.~{Kramer}.
\newblock {A characteristic observable signature of preferred-frame effects in
  relativistic binary pulsars}.
\newblock {\em \mnras}, 380:455--465, 2007.

\bibitem{wil93}
C.~M. {Will}.
\newblock {\em {Theory and Experiment in Gravitational Physics}}.
\newblock Cambridge University Press, 1993.

\bibitem{wil06} 
C.~M.~Will.  
\newblock {The Confrontation between
  General Relativity and Experiment}.  
\newblock {\em Living
  Rev. Relativity}, 9:3, 2006. 
[URL (cited on 2013-02-26):
  http://www.livingreviews.org/lrr-2006-3]

\bibitem{wn72}
C.~M. {Will} and K.~{Nordtvedt}.
\newblock {Conservation Laws and Preferred Frames in Relativistic Gravity. I.
  Preferred-Frame Theories and an Extended PPN Formalism}.
\newblock {\em \apj}, 177:757, 1972.

\bibitem{yhc+07}
X.~P.~You, G.~Hobbs, W.~A.~Coles, \etal.
\newblock {Dispersion measure variations and their effect on precision
pulsar timing}.
\newblock {\em \mnras}, 378:493, 2007.

\end{thebibliography}
\end{document}